\documentclass[fleqn,usenatbib]{mnras}

\usepackage[T1]{fontenc}
\usepackage{ae,aecompl}

\usepackage{graphicx}	
\usepackage{amsmath}	
\usepackage{amssymb}	

\usepackage{etoolbox}
\makeatletter
\patchcmd\@combinedblfloats{\box\@outputbox}{\unvbox\@outputbox}{}{%
  \errmessage{\noexpand\@combinedblfloats could not be patched}%
}%
\makeatother

\newcommand{\be}{\begin{equation}}
\newcommand{\ee}{\end{equation}}
\newcommand{\bea}{\begin{eqnarray}}
\newcommand{\eea}{\end{eqnarray}}

\title[Optical Substructure of SZ and X-ray Clusters]{Optical Substructure and BCG Offsets of Sunyaev-Zel'dovich and X-ray Selected Galaxy Clusters}

\author[P. A. A. Lopes et al.]{
Paulo A. A. Lopes$^{1,2}$\thanks{E-mail: plopes@astro.ufrj.br},
M. Trevisan$^{3,4,2}$,
T. F. Lagan\'a$^{5}$,
F. Durret$^{2}$,
A. L. B. Ribeiro$^{6}$,
\newauthor
S. B. Rembold$^{7}$
\\
$^{1}$Observat\'orio do Valongo, Universidade Federal do Rio de Janeiro, Ladeira do Pedro Ant\^onio 43, Rio de Janeiro, RJ, 20080-090, Brazil\\
$^{2}$Institut d'Astrophysique de Paris (UMR 7095: CNRS, Sorbonne Universit\'e), 98 bis Bd Arago, F-75014 Paris, France\\
$^{3}$Universidade Federal do Rio Grande do Sul -- Departamento de Astronomia -- 91501-970, Porto Alegre-RS, Brazil\\
$^{4}$GEPI, Observatoire de Paris, PSL University, CNRS, Place Jules Janssen 92195, Meudon, France\\
$^{5}$N\'ucleo de Astrof\'isica Te\'orica, Universidade Cruzeiro do Sul, Rua Galv\~ao Bueno 868, Liberdade, CEP: 01506-000, S\~ao Paulo,SP, Brazil\\
$^{6}$Laborat\'orio de Astrof\'isica Te\'orica e Observacional -- Departamento de Ci\^encias Exatas e Tecnol\'ogicas --Universidade Estadual de Santa Cruz,\\ 45650-000, Ilh\'eus, BA, Brazil\\
$^{7}$Universidade Federal de Santa Maria -- 97105-900, Santa Maria-RS, Brazil\\
}

\date{Accepted 2018 May 23. Received 2018 May 19; in original form 2018 April 19}

\pubyear{2018}

\begin{document}
\label{firstpage}
\pagerange{\pageref{firstpage}--\pageref{lastpage}}
\maketitle

\begin{abstract}
We used optical imaging and spectroscopic data to derive substructure estimates for local Universe ($z < 0.11$) galaxy clusters from two different samples. The first was selected through the Sunyaev-Zel'dovich (SZ) effect by the Planck satellite and the second is an X-ray selected sample. In agreement to X-ray substructure estimates we found that the SZ systems have a larger fraction of substructure than the X-ray clusters. We have also found evidence that the higher mass regime of the SZ clusters, compared to the X-ray sample, explains the larger fraction of disturbed objects in the Planck data. Although we detect a redshift evolution in the substructure fraction, it is not sufficient to explain the different results between the higher-z SZ sample and the X-ray one. We have also verified a good agreement ($\sim$60$\%$) between the optical and X-ray substructure estimates. However, the best level of agreement is given by the substructure classification given by measures based on the brightest cluster galaxy (BCG), either the BCG$-$X-ray centroid offset, or the magnitude gap between the first and second BCGs. We advocate the use of those two parameters as the most reliable and cheap way to assess cluster dynamical state. We recommend an offset cut of $\sim$0.01$\times$R$_{500}$ to separate relaxed and disturbed clusters. Regarding the magnitude gap the separation can be done at $\Delta m_{12} = 1.0$. The central galaxy paradigm (CGP) may not be valid for $\sim$20$\%$ of relaxed massive clusters. This fraction increases to $\sim$60$\%$ for disturbed systems.
\end{abstract}




\section{Introduction}

Galaxy clusters are the largest collapsed structures in the Universe, representing highly biased peaks of the dark matter large scale structure (LSS). The cluster mass function and its time evolution is a key cosmological probe, as the variation  in the cluster abundance over cosmic time strongly depends on many cosmological parameters, such as $\sigma_8$, $\Omega_m$ and $\Omega_{\lambda}$ \citep{eke98,bah03,roz10,deH16,man14,all11,sch17,pla16}.

To use cluster number counting to constrain cosmological parameters, the cluster total mass is primordial. Masses can be estimated using X-ray data under the assumption of hydrostatic equilibrium, through measurements of the velocity dispersion of the cluster galaxies (under the assumption of virial equilibrium) and by means of weak and strong lensing. Deriving masses through
one of these methods is impractical for large samples, at any redshift, as they are observationally very expensive. Hence, the common approach is to use robust
scaling relations that relate total masses with an easily observed quantity 
such as X-ray luminosity (L$_X$), richness (N$_{gal}$), or optical luminosity (L$_{opt}$) \citep{lop09b,roz10}.

To construct such relations, the wavelength of cluster selection plays a significant role. Because different physical mechanisms are responsible for radiation production in different wavelength ranges, clusters of galaxies detected using different techniques do not constitute a homogeneous population.
These systematic differences between samples, which can include the degree of cluster virialization, impact both the slope and the scatter of cluster scaling relations. Therefore an accurate knowledge of biases related to the wavelength used for the identification and characterization of clusters of galaxies is a necessary step to allow the use of these systems as cosmological tools.
The approach to address these biases is the investigation of complementary properties of clusters selected in other wavelength regimes \citep{don02,lop06,dai07,gal09}.

This issue has been highlighted in the last few years with the availability of cluster samples selected through the SZ effect by the Planck satellite. Due to the different dependence of the SZ effect and X-ray emission on the gas density, there is currently a
debate regarding whether the two experiments are detecting the same population of galaxy clusters.
In particular, it has been shown that the Planck catalogs contain a smaller fraction of cool-core (hereafter CC) or relaxed systems in comparison to flux limited X-ray samples \citep{and17,plc11,ros17}. The presence of substructure is a clear sign of incomplete relaxation in a cluster, that can also impact the scatter of cluster scaling relations. Even the overabundance of CC clusters in X-ray flux limited samples leads to an increased scatter of the mass-luminosity relation \citep{cho12}.

Estimates of the fraction of cluster substructure vary from $\sim$20 to $\sim$80\%. One of the main reasons for this large scatter is the method employed for substructure detection \citep{kol01}, but the choice of centroid, radius, magnitude range and, of course, the wavelength of observation are also important.
\citet{lop06} applied four substructure tests to photometric data of 10190 optically selected clusters finding that the fraction of disturbed systems varies between 13 and 45\%. \citet{wen13} also used photometric data to develop a method to estimate cluster substructure. After applying it to 2092 rich clusters they found that 28\% exhibit signs of substructure. Using photometric and spectroscopic information \citet{lop09a,lop09b} found that 23\% of clusters are not relaxed according to the $\Delta$ test \citep{dre88}. They also found that the exclusion of systems with substructure has no impact in cluster scaling relations. Regarding the comparison of the CC fraction in X-ray and SZ samples, it has been found that the former contain a larger fraction of CC systems. For instance, \citet{ros17} found 59 {\it vs} 29\% of CC systems in X-ray and SZ samples, respectively. \citet{and17} found similar results after employing four indicators of the CC state. \citet{lov17} corroborate those results using eight morphological parameters applied to SZ and X-ray clusters.

Until today this comparison of the fraction of relaxed clusters in SZ and X-rays has not been done using optical substructure estimates. This is one of the main goals of the current paper, for which we used the SZ and X-ray samples presented by \citet{and17}. We also compare the performance of different  estimators of the cluster dynamical state: the four X-ray CC measures from \citet{and17}, plus the six optical estimates we adopt in this paper (two of which are based on the properties of the brightest cluster galaxies, BCGs). We show here that the most reliable and cheap way to assess a cluster dynamical state is through the BCG offset and this result has a promising impact on studies that rely on large samples of galaxy clusters.

This paper is organized as follows: in $\S$2 we describe the cluster samples, as well as the galaxy redshift surveys used in this work and the derivation of cluster properties from their galaxy distribution ($\sigma_P$, R$_{500}$, M$_{500}$, R$_{200}$, and M$_{200}$). $\S$3 is devoted to  describe the selection of the first and second brightest cluster galaxies, while $\S$4 describes the optical substructure tests we employed. Our main results are presented in $\S$5 and further discussed in $\S$6. In $\S$7 we draw our conclusions. The cosmology assumed in this work is $\Omega_{\rm m}= $ 0.3, $\Omega_{\lambda}= $ 0.7, and H$_0 = 100$ $\rm h$ $\rm km$ $s^{-1}$ Mpc$^{-1}$, with $\rm h$ set to 0.7.

\section{Data}

\subsection{Cluster Sample} 
\label{cls-sample} 

This work is based on the two samples presented by \citet{and17}. The first is based on the Planck Early Sunyaev-Zel'dovich (ESZ) sample of 189 SZ clusters, containing 164 clusters at $z < 0.35$. These clusters have been followed up in X-rays by the {\it Chandra-Planck Legacy Program for Massive Clusters of Galaxies}\footnote{hea-www.cfa.harvard.edu/CHANDRA\_PLANCK\_CLUSTERS/}. The second data set is composed of 100 clusters from a flux-limited X-ray sample for which Chandra data are also available. Most ($\sim $90\%) of the clusters in this data set are located below $z = 0.1$, with the remaining objects spread up to $z \sim 0.2$. The Planck ESZ and X-ray samples have 49 clusters in common.

Here we want to estimate optical substructure for those objects, using three dimensional (RA, DEC, $z$) data for their members. We have drawn these data from different surveys. In the northern hemisphere our data come from the Sloan Digital Sky Survey (SDSS), while in the south we use the 2dF Galaxy Redshift Survey (2dFGRS) and the 6dF Galaxy Survey (6dFGS). Our data set is then limited to $z = 0.11$ as we respect the redshift limits of these three surveys ($z = 0.10$ for SDSS, $z = 0.11$ for the 2dF and $z = 0.055$ for the 6dF). As the completeness of these surveys is affected by issues like fiber collision, we have also gathered additional redshifts from the NASA/IPAC Extragalactic Database (NED).

We start by selecting all clusters, from the two data sets of \citet{and17}, with a redshift value smaller than each of the surveys above and falling within the footprint of those surveys. From the 164 Planck ESZ objects we have found 22 clusters within SDSS, 2 in the 2dF region and 18 in the 6dF footprint. The combined optical data for the ESZ has 40 clusters. For each of the common clusters among the three data sets we keep the one with more galaxies available. Out of the 100 X-ray systems we found 33 objects in SDSS, 6 in the 2dF and 27 in the 6dF survey. The combined X-ray sample after accounting for clusters selected twice has 62 systems. Note that we have fewer ESZ clusters in comparison to the X-ray sample (the opposite of the original samples) as most of the 164 ESZ clusters are above $z = 0.1$, as seen in Fig. 1 of \citet{and17}. The opposite is true for the X-ray sample. Our combined ESZ and X-ray sample has 72 clusters, with 30 objects in common to the two data sets. Among the 72 clusters there are 10 secondary subclusters identified in X-rays. We kept those separately as done by \citet{and17}.

Table~\ref{tab:cls_props} lists the main characteristics of the seventy two clusters of this work. The derivation of these cluster properties is described in $\S$ \ref{cls_prop}. The cluster name is in column 1; coordinates are shown in columns 2 and 3; redshift in 4; velocity dispersion is in column 5; while the characteristic radii and masses (R$_{500}$, M$_{500}$, R$_{200}$, and M$_{200}$) are in columns 6 to 9.

\begin{figure*}
\begin{center}
\leavevmode
\includegraphics[width=6.5in]{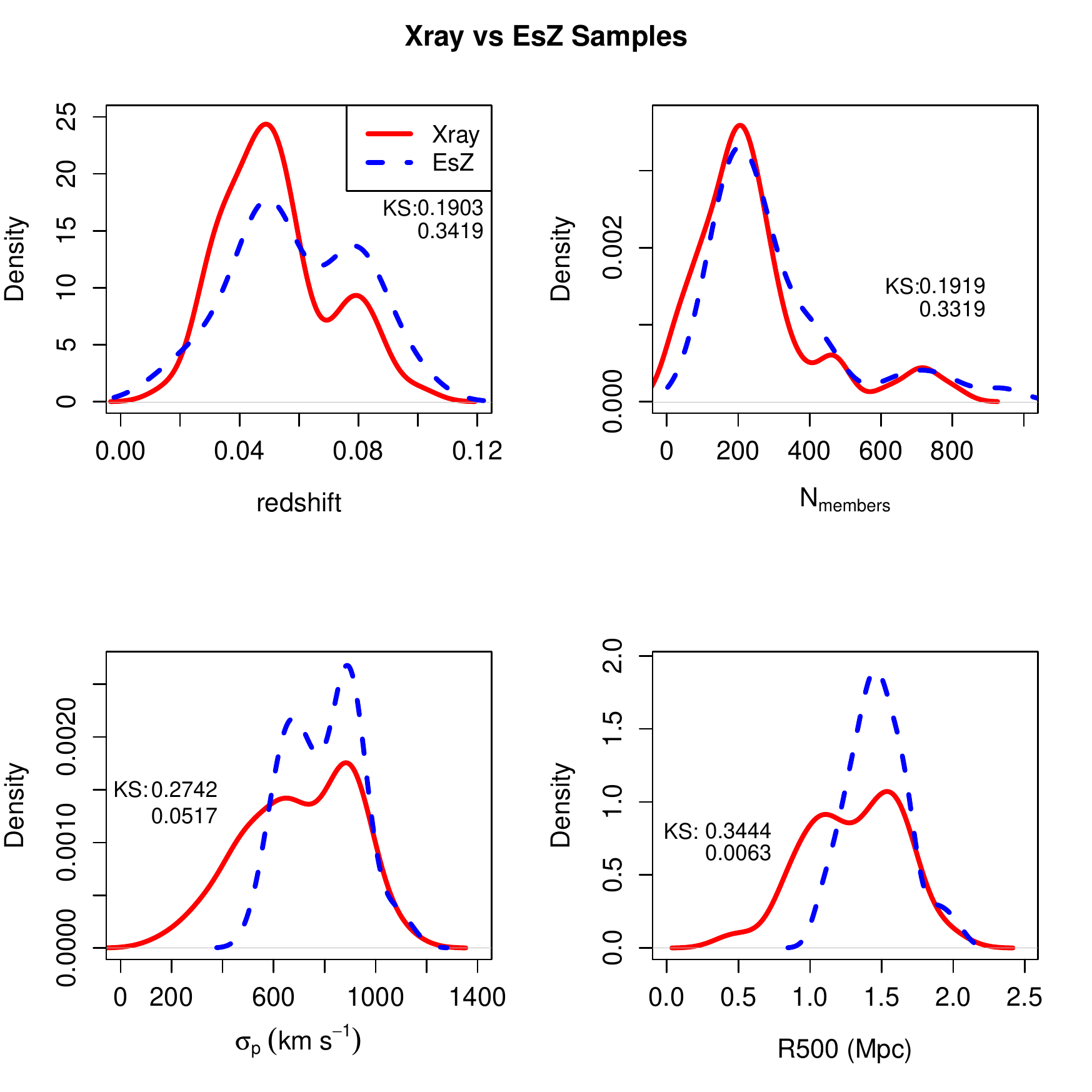}
\end{center}
\caption{Optical properties of the two samples used in this work. The X-ray sample is displayed by the red solid line and the ESZ is the blue dashed line. On the top left we compare the redshift distributions, while the total number of members per cluster is in the top right. In the bottom panels we show the velocity dispersion (left) and R$_{500}$ distributions (right). In all panels we also show the value of the KS test and $p-$value from the comparison of the samples.}
\label{fig:cls_opt_props}
\end{figure*}

In Fig.~\ref{fig:cls_opt_props} we display some of the optical properties (see $\S$ \ref{cls_prop}) of the two samples (ESZ and X-ray). The distributions of redshift (left) and number of members (right) are in the top panels, while the velocity dispersion is on the left bottom panel and R$_{500}$ is in the bottom right. The Kolmogorov-Smirnov (KS) statistics between the two samples and the respective $p-$value of the test are also listed. From that we conclude that the distributions of redshift and number of members are not significantly different between the two samples. However, their mass distributions (indicated by their physical radius and velocity dispersions) are different. Although, the current work is limited to lower$-z$ the ESZ sample has more massive clusters (as in the original samples).

\subsection{Sloan Digital Sky Survey} \label{sdss}

The Sloan Digital Sky Survey (SDSS) is an imaging and spectroscopic survey that began in 2000, progressing in different phases. Currently, it is on its 4th phase. SDSS obtained deep multi-band images for one third of the sky and acquired spectra for more than three million celestial objects. The SDSS data are public with periodic releases to the astronomical community. Currently SDSS is on its data release 14 (DR14).

In order to have the largest number of redshifts we initially planned to use the SDSS DR14 data. However, due to the changes in the photometric pipeline after the DR8 \citep{dr8sdss2011} some bright galaxies have their fluxes underestimated. We noticed this when selecting the brightest cluster galaxies (BCGs) of our sample. Hence, we decided to use all galaxies with spectra in SDSS DR14 or DR7, keeping the DR7 magnitude whenever available. Otherwise we used the DR14 magnitude. Nonetheless, magnitudes are only used in our work for studying the first- and second-ranked galaxies in the cluster, which we refer to, for simplicity, as ``first BCG'' and ``second BCG'', respectively. For each cluster from \citet{and17} we searched for galaxies in SDSS within 5 Mpc of the X-ray centroid.

\subsection{2dF Galaxy Redshift Survey} \label{2df}

The 2dF Galaxy Redshift Survey (2dFGRS) is a spectroscopic survey conducted in the southern hemisphere at the Anglo-Australian Observatory. The survey and its database are described in \citet{col01}, with its final release happening in 2003, with a total number of 245591 unique objects with spectra, mainly galaxies. The survey was designed to obtain spectra for objects brighter than an extinction-corrected magnitude of $b_J = 19.45$. As above we searched for galaxies for every cluster using an aperture of 5 Mpc. We converted the SuperCOSMOS $r_F$ magnitudes to the SDSS photometric system using equation 1 from \citet{pea16}.

\subsection{6dF Galaxy Survey} \label{6df}

The 6dF Galaxy Survey (6dFGS) mapped nearly half of the nearby Universe over six years ($\sim 17000$ square degrees of the southern sky), yielding a new catalogue of 125071 galaxies with median redshift $z = 0.053$. The magnitude limit of the sample is $b_J = 16.75$ or $r_F = 15.60$. Its final data release was available on 2009 \citep{jon09}. We used all the 125071 galaxies from the 6dF to search for galaxies within 5 Mpc from the X-ray centers. As above, the SuperCOSMOS $r_F$ magnitudes were transformed to the SDSS photometric system through the use of equation 1 from \citet{pea16}.

\subsection{NASA/IPAC Extragalactic Database redshifts} \label{NED}

The NASA/IPAC Extragalactic Database (NED\footnote{http://ned.ipac.caltech.edu}) database is a multiwavelength collection of information on extragalactic objects, being actively updated with different sky surveys and results from research publications. We selected all galaxies in NED with a reliable spectroscopic redshift within 5 Mpc from each cluster in our sample. For this selection we considered the quality code on the redshifts. We used the galaxies with the code listed as ``blank" (``usually a reliable spectroscopic value'') or ``SPEC'' (``an explicitly declared spectroscopic value''). We verified the magnitude distributions of the galaxy samples after including the NED redshifts and found we reach at least the same limits as the original data sets (SDSS, 2dF and 6dF).

\begin{figure*}
\begin{center}
\leavevmode
\includegraphics[width=6.5in]{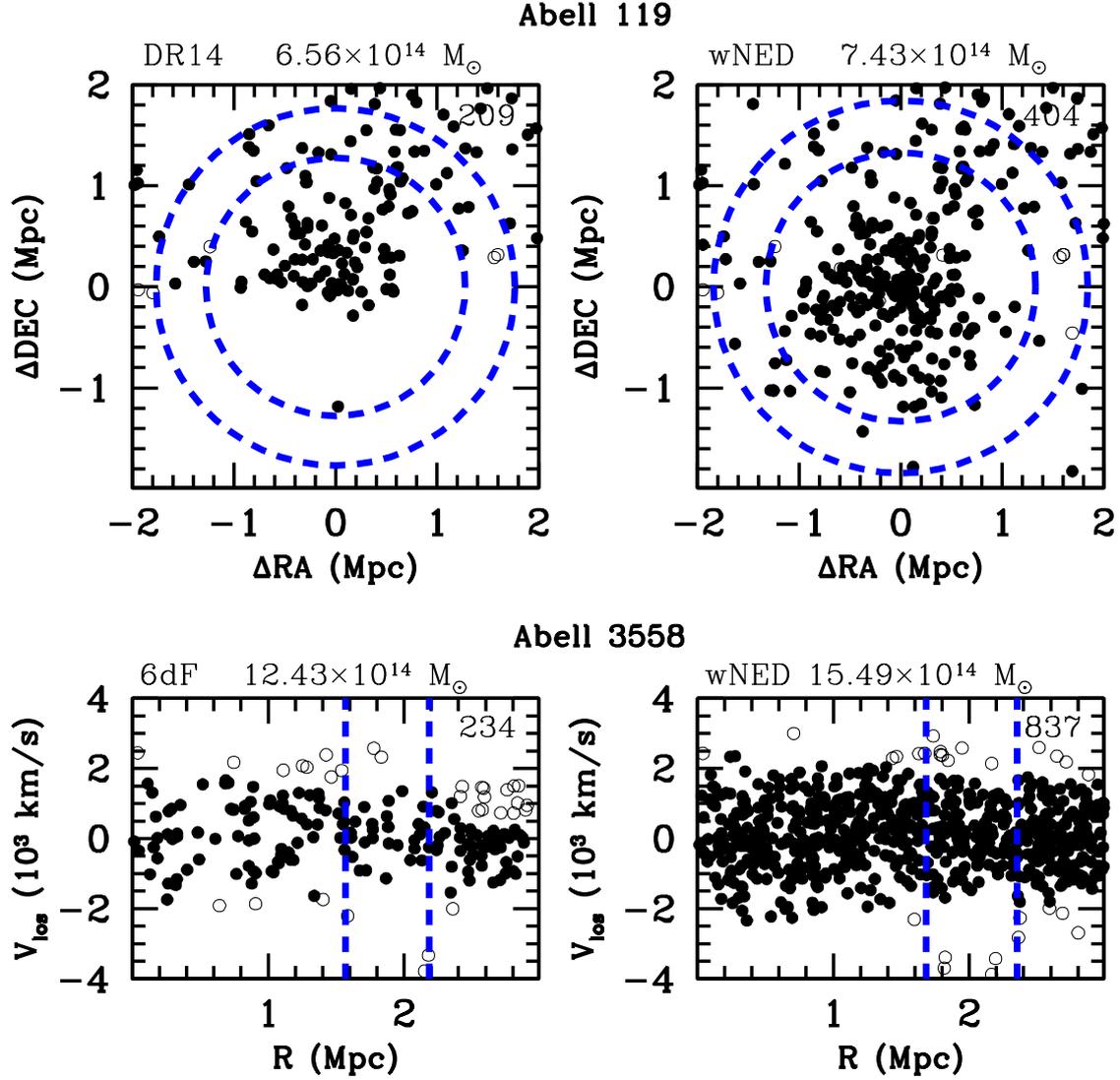}
\end{center}
\caption{Example of two clusters for which the NED redshifts represented an important step towards better mass and substructure estimates. In the top panels we show the projected distribution of galaxies in the cluster Abell 119. The SDSS DR14 distribution is on the left and the one with NED galaxies on the right. The bottom panels display the phase-space distribution of galaxies in the cluster Abell 3558. On the left we have the 6dF galaxies, while on the right the distribution with NED objects is shown. On all panels, member galaxies are shown with filled symbols, while interlopers are displayed by open circles. The estimated mass and total number of galaxies available (including interlopers) are given on the top of each panel. On all panels, the dashed lines indicate R$_{500}$ and R$_{200}$.}
\label{fig:cls_example_NED_gals}
\end{figure*}

The use of NED galaxies greatly improved the number of redshifts available for the clusters in our sample. For northern clusters (SDSS footprint) the median increment in the number of redshifts available was of $\sim 7\%$. This is in good agreement with the expected loss due to the fiber collisions in SDSS. For the 6dF sample in the south the number of galaxies available increased by $\gtrsim 3$ times. In Fig.~\ref{fig:cls_example_NED_gals} we show an example of two clusters for which the extra galaxies from NED represented an important addition to our analysis. This is easily seem in the comparison between the left and right panels (before and after inclusion of NED redshifts). 

\subsection{Galaxy Cluster Properties}\label{cls_prop}

For each cluster we adopted the X-ray centroid listed by \citet{and17}. However, as we have the spectroscopic redshifts of galaxies in the region of all clusters we recomputed the cluster redshift. We do so after applying the gap technique described in \cite{kat96}, using a density gap \citep{ada98, lop07, lop09a} that scales with the number of galaxies available. The gap technique is used to identify groups in redshift space. We applied it to all galaxies within 0.50 h$^{-1}$ Mpc of the cluster centre. Within such a small aperture it is common to identify only one significant group. However, if more than one is found we keep the one that is closest to the cluster X-ray center. The cluster redshift is then given by the biweight estimate \citep{bee90} of the galaxy redshifts of the chosen group. As a byproduct we also obtain the velocity limits ($v_{lo}$ and $v_{hi}$) of the cluster within this radius. These values are used as input for the code employed to reject interlopers and derive a final list of cluster members.

The procedure adopted for the membership selection is the `shifting gapper' technique \citep{fad96}, described in detail at \citet{lop09a}. It is based on the application of the gap technique in radial bins from the cluster centre. We adopt a bin size of 0.42 h$^{-1}$ Mpc (0.60 Mpc for h $= 0.7$) or larger if less than 15 galaxies are selected. In every bin we eliminate galaxies not associated to the main body of the cluster.
The procedure is repeated until no more galaxies are rejected as interlopers. One great advantage of the method is to make no hypothesis about the dynamical status of the cluster. When running this procedure we consider all galaxies within 2.5 h$^{-1}$ Mpc (3.57 Mpc for h $= 0.7$) from the cluster center and with $|cz - cz_{cluster}| \le 4000~{\rm km~s}^{-1}$.

Next we perform a virial analysis to obtain estimates of the velocity dispersion, the physical radius and mass ($\sigma_P$, $R_{500}$, $R_{200}$, $M_{500}$ and $M_{200}$). We derive a robust velocity dispersion estimate ($\sigma_P$) using the gapper or biweight estimator, depending if $< 15$ (gapper) or $\ge 15$ (biweight) galaxies are available \citep{bee90}. The velocity dispersion is also corrected for velocity errors. We then obtain an estimate of the projected `virial radius' and a first estimate of the virial mass \citep{gir98}. Assuming a NFW profile \citep{nav97} and applying the surface pressure term correction to the mass estimate, we obtain final estimates for the mass ($M_{500}$ and $M_{200}$), as well as for the physical radius ($R_{500}$ and $R_{200}$).

\section{Brightest cluster galaxy identification}

To select the BCGs we adopt a similar approach as the one described in \citet{lav16}. We have also considered a maximum radius of 0.5$\times$R$_{500}$ to identify the BCG among all the previously selected member galaxies. The main differences here are the use of the $r-$band and the fact we have a robust membership selection. As we are restricted to low$-z$ clusters, for which a large number of spectroscopic redshifts are available, we can apply the `shifting gapper' technique described in $\S$ \ref{cls_prop}. To obtain absolute magnitudes in the $r-$band we employ the formula: $M_r = m_r - DM - kcorr - Qz$, where DM is the distance modulus (using the galaxy redshift), $kcorr$ is the k$-$correction and $Qz$ ($Q = -1.4$, \citealt{yee99}) is a mild evolutionary correction applied to the magnitudes. For the SDSS data we use the k$-$correction derived through a template
fitting procedure, while for galaxies in the 2dF and 6dF regions we used a k$-$correction typical of elliptical (E) galaxies \citep{lop09a,lop09b} obtained through the convolution of an E spectral energy distribution with the SDSS $r$ filter.

The approach above normally results in a reliable BCG selection as the galaxy presenting the highest r-band luminosity. However, in a few cases the BCG has not been spectroscopically observed or the available magnitude is wrong (fainter than the expected value). Hence, for the galaxies in the SDSS footprint we verified if a brighter galaxy without a spectrum could be the BCG. A visual inspection was also performed (by PAAL) for the central regions of all clusters. In the north we used the {\it navigate}\footnote{http://skyserver.sdss.org/dr14/en/tools/chart/navi.aspx} tool available in the SDSS sky server. For clusters in the south (2dF and 6dF areas) the visual inspection was based on the {\it ESASky 2.0} portal\footnote{http://sky.esa.int}, from which we checked the DSS2 color images.

Differently from \citet{lav16} we used the same radius 0.5$\times$R$_{500}$ for the selection of the second brightest cluster galaxy. They modified the search radius to R$_{500}$, which can lead to wrong selection due to larger background/foreground contamination. That is not a problem in our case, as we use only spectroscopically selected members. However, a search radius larger than the one used for the first BCG selection may also result in the selection of the first-ranked galaxy of a nearby system, {\it e.g.} in case of merging clusters. As many Planck clusters show substructure and some of them are merging clusters we may find systems separated by small distances ($\lesssim $ R$_{500}$). Hence, the use of the same radius to select the BCG and second brightest cluster galaxies seems more appropriate.

\section{Substructure estimates}

Based on the optical data described above we derived six estimates of the dynamical stage of the clusters. The first three methods listed below are fully described in \citet{pin96}, who evaluated the performance of thirty-one statistical tests. The three tests we employed are: the Dressler \& Shectman (DS or $\Delta$), the symmetry ($\beta$) and the Lee3D statistics. In one dimension, using galaxy velocities, we used the Anderson-Darling (AD) statistic. The other two diagnostics are based on the offset between the BCG position and the X-ray centroid, as well as the magnitude difference of the first and second brightest cluster galaxies.

We apply the substructure tests for all galaxies inside an aperture of radius R$_{500}$ around the cluster center. The AD requires a minimum number of eight galaxies. This is true for all systems in our sample, except for one object. For this cluster we enlarged the radius by 10\% (less than the R$_{500}$ error), only to run the AD test, to guarantee the minimum number. We chose to work with this aperture (R$_{500}$) to be in agreement with the one used for investigating substructures in X-rays with Chandra data \citep{and17}. 

The significance level of the three substructure tests from \citet{pin96} is determined through Monte Carlo simulations, for which we have five-hundred realizations. We compute the number of Monte Carlo simulations which resulted in a larger value for a given substructure statistics than the real data. Then we divide this number by the number of realizations. We adopt a significance threshold of 5$\%$. This means that a substructure estimate is considered as significant if at most twenty-five simulated data sets have substructure statistics higher than the observations. Further details can be found in \citet{pin96} and \citet{lop06}. In Table \ref{tab:sub_props} we list the values of the three substructure tests from \citet{pin96} and their significance levels. We also list the AD statistic and its associated $p-$value, as well as the BCG coordinates, offsets and magnitude gaps. A brief description of the six optical substructure tests is given below. 

Regarding the X-ray estimates, the four parameters used by \citet{and17} to identify cool-core clusters are: (i) the concentration parameter (C$_{SB}$), defined as the ratio of the integrated emissivity profile within 0.15$\times$R$_{500}$ to that within R$_{500}$; (ii) another concentration parameter (C$_{SB4}$), which is the ratio of the integrated
emissivity profile within 40 kpc to that within 400 kpc; (iii) the cuspiness of the gas density profile ($\delta$), the negative of the logarithmic derivative of the gas density with respect to the radius, measured at 0.04$\times$R$_{500}$; and (iv) the central gas density (n$_{core}$), measured at 0.01$\times$R$_{500}$. For further details we refer the reader to the work of \citet{and17}.

\subsection{Optical substructure}

\subsubsection{The Dressler \& Shectman (DS) test}

The first substructure test we employed is the three dimensional DS or $\Delta$ test \citep{dre88}. The algorithm computes the mean velocity and standard deviation ($\sigma$) of each galaxy and its N$_{nn}$ nearest neighbors, where N$_{nn} = $ N$^{1/2}$ and N is the number of galaxies in the cluster region. These local values are compared to the global mean and $\sigma$ (obtained with all galaxies). A deviation from the global value is given by equation \ref{eq:DS}. Substructure is estimated with the cumulative deviation $\Delta$ ($\sum \delta_i$). Objects with no substructure have $\Delta \sim$ N.

\bea
{\rm \delta_i^2} = \left( \frac{N_{nn}+1}{\sigma^2} \right)[(\bar{v}_{local}-\bar{v})^2 + (\sigma_{local} - \sigma)^2]. \,\,\,
\label{eq:DS}
\eea

\subsubsection{The Symmetry test}

The two-dimensional Symmetry or $\beta$ test was introduced by \citet{wes88}. This test searches for significant deviations from mirror symmetry about the cluster center. For every galaxy ``i'', a local density estimate d$_i$ is obtained from the mean distance to the N$^{1/2}$ nearest neighbors. The local density, d$_{\circ}$, for a point ``$\circ$'' diametrical to a galaxy ``i'' is also obtained. For a symmetric galaxy distribution the two estimates should, on average, be approximately equal. This is not the case for clumpy distributions. The asymmetry for a given galaxy ``i'' is given by

\bea
{\rm \beta_i} =  \rmn{log}\biggl(\frac{d_{\circ}}{d_i}\biggr), \,\,\,
\label{eq:beta}
\eea

The $\beta$-statistic is defined as the average value $<$$\beta_i$$>$ over all galaxies. For a symmetric distribution $<\beta> \approx 0$, but values of $<$$\beta$$>$ significantly larger than 0 indicate asymmetries. 

\subsubsection{The Lee3D test}

The Lee statistic \citep{lee79} is a test of bimodality in distributions, based on a maximum likelihood technique to separate a data set of two or more dimensions into two parts. 

In two dimensions the algorithm begins by projecting the N points onto a line, making the angle $\phi$ relative to a second line. The first line is then rotated in small steps in the range $0^{\circ} < \phi < 180 ^{\circ}$. For each different orientation the points assume a new coordinate as they are projected onto the line. Next, a search for the best partition into a ``left'' and ``right'' clump is performed. For all $N-1$ partitions $\sigma_l$, $\sigma_r$ and $\sigma_T$ are computed for the left, right, and total samples, respectively. The Lee statistic (L) is a function of those values. In three dimensions the velocity dispersion is used as a weight in the computation of L. The main advantage of the Lee 3D test is the insensitivity to non-substructure that may appear as substructure to other tests (such as elongation).

\subsubsection{The Anderson-Darling test}

The Anderson-Darling (AD) test is a normality statistical test, based on the comparison of the empirical distribution function (EDF) and the ideal case of a normal distribution (for further details see \citealt{hou09, rob18}). The test does not require binning or graphical analysis. In the recent years it has been more commonly used in astronomy \citep{hou09, rib10, rib11, rib13a, rib13b, kra13, rob18}.

We apply the AD test to the velocity distribution of galaxies in the clusters. For relaxed systems the galaxy distribution is expected to be well represented by a normal, while perturbed systems should be characterized by larger deviations from normality.

For the current work we follow the simple prescription given by \citet{rob18} to classify Gaussian (G) and non-gaussian (NG), or relaxed or not, clusters. They chose a critical 
p$-$value of 0.10 from the AD test. Relaxed (G) clusters have p$-$values $p_{AD} > 0.10$, while disturbed (NG) systems have $p_{AD} \le 0.10$. They mention that their results are not sensitive to the precise p$-$value choice around this critical value.

\subsubsection{The BCG X-ray centroid offset}\label{bcg_off}

The X-ray emission peak is a good representation of the bottom of the gravitational potential of galaxy clusters, being a good indicator of the center of the X-ray distribution. It is also well known that BCGs should, in general, be located very close to the cluster center \citep{jon84,pos95,lin04, lin17}. Hence, the offset between the X-ray center and the BCG position can be used as an indication of the dynamical state of a cluster. We expect nearly zero positional offsets for relaxed clusters, but non-negligible offsets for disturbed systems.

\begin{figure}
\begin{center}
\leavevmode
\includegraphics[width=3.5in]{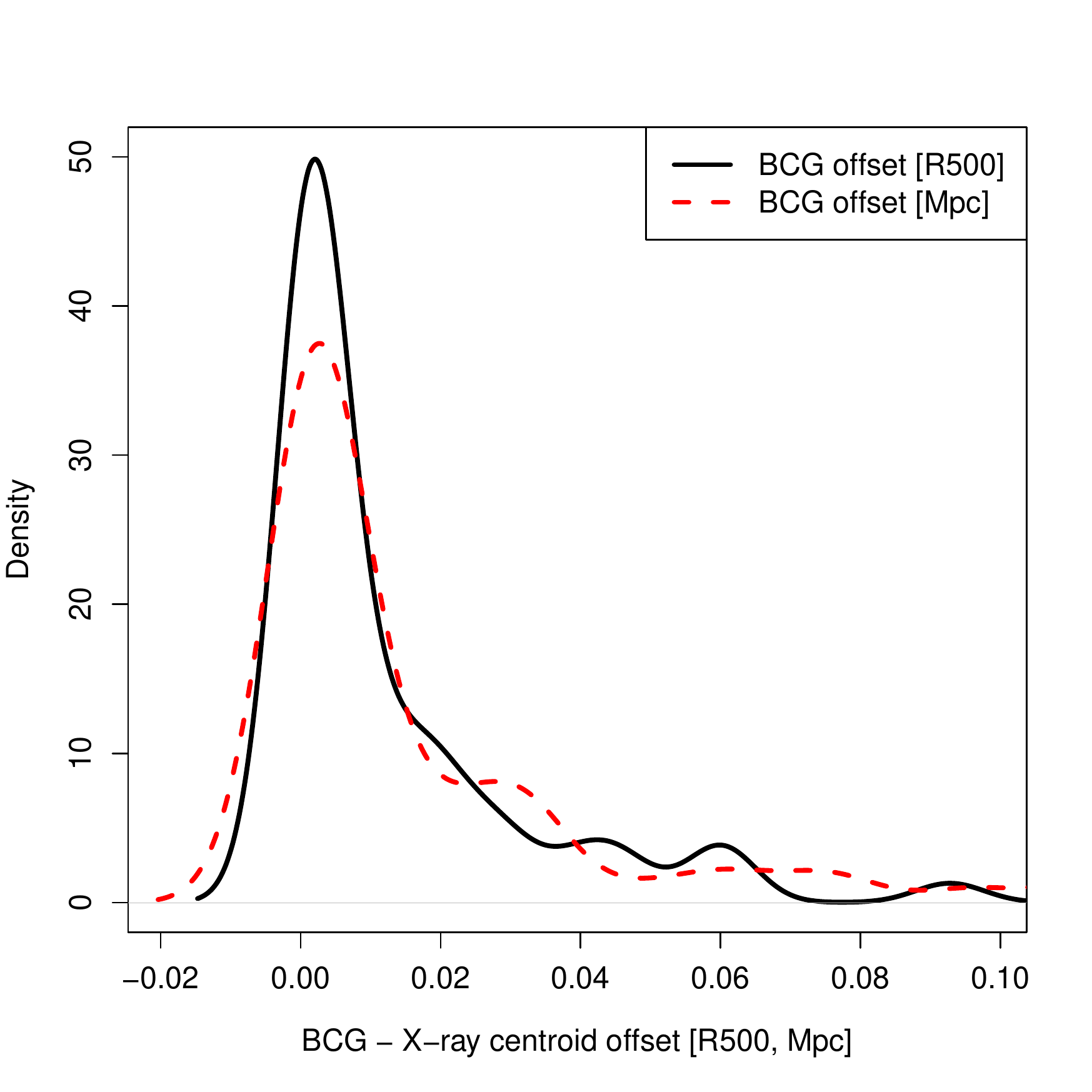}
\end{center}
\caption{BCG$-$X-ray offset distribution. The solid black line displays the offsets in units of R500, while the red dashed line is in Mpc.}
\label{fig:bcg_offset}
\end{figure}

Initially, in agreement with \citet{lav16}, we adopted an offset between the BCG and the X-ray centroid of 0.05$\times$R$_{500}$ to separate relaxed and non-relaxed systems. However, we decided to search for an optimal break value of the BCG to X-ray centroid offset, finding that a more rigorous criterion works better (see details in $\S$ \ref{comp_sub}). We then decided to classify clusters as relaxed if the BCG$-$X-ray offsets are $<$ 0.01$\times$R$_{500}$. Disturbed objects are those with offsets $\ge$ 0.01$\times$R$_{500}$. Fig.~\ref{fig:bcg_offset} shows the BCG$-$X-ray offset distribution, in units of R500 (solid line) and in Mpc (dashed line).

This criterion roughly corresponds to an offset of $\sim$10 kpc. Taking into account the Chandra angular resolution of $0.5''$ and a $0.16''$ astrometric  error for point sources in the  Chandra Source Catalog (CSC, \citealt{rot11}), it is reasonable to conservatively assume an error $< 1''$ for extended sources. At the maximum redshift of our sample ($z = 0.11$) the angular scale is 2.006 kpc/$''$. Hence, using the Chandra X-ray peak measurements from \cite{and17} the offset we want to probe (0.01$\times$R$_{500}$ or $\sim$10 kpc) is at least five times the angular error in the X-ray centroid. At redshift one this offset would still be larger than the Chandra precision.

\subsubsection{The magnitude difference between the first and second BCGs}

The central location of BCGs also favours their growth through the accretion of minor galaxies that are driven to central parts of clusters. Since dynamical-friction timescales are inversely proportional to the galaxy mass, more massive satellites tend to merge with the central galaxy before the less massive ones. As a result the BCG tends to increase its luminosity difference over time relative to its neighbours.  This is measured by the ``magnitude gap'' between the BCG and other cluster galaxies, traditionally the second brightest cluster galaxy. Thus, the luminosity gap $\Delta$m$_{12}$ provides a measure of galaxy, but also cluster, evolution \citep{tre77,pos95,lav16,gol17}. 

In the current work we try to assess the possibility of using the magnitude gap $\Delta$m$_{12}$ as an indicator of the cluster dynamical status. We found in $\S$ \ref{results} that at a value of $\Delta$m$_{12} = 1.0$ it is possible to separate relaxed and perturbed clusters, the latter having smaller gaps.

\section{Results} \label{results}

\subsection{Fraction of Relaxed Clusters} \label{frac_rel}

We summarize our main results regarding the fraction of relaxed clusters identified by the different substructure tests in Table \ref{tab:relaxed_frac}. The percentage of clusters classified as relaxed according to each test is listed for the ESZ and X-ray samples. In the last column we indicate the number of clusters in each sample. In the last two lines we also show the results with optimal values for the X-ray classification. Our motivation for that  comes from the analysis of \citet{lov17}, who compared the performance of different X-ray morphological estimators. Some other works doing that in X-rays are \citet{ras13}, who used X-ray simulations and \citet{don16} who worked with X-ray, but also SZ and mass maps. In particular, using eight X-ray morphological parameters to classify clusters, \citet{lov17} find that the distributions of all the parameters show an overlap between relaxed and disturbed clusters. Even for the parameters with little overlap (centroid-shift, concentration, and power ratio) there is no clear cut between the two populations. Hence, instead of using a simple value to classify regular and disturbed systems they provide limits (or cuts) for each parameter to construct samples with high completeness of relaxed/disturbed clusters. These samples may not have high purity, so that \citet{lov17} also provide the best cuts to build samples of relaxed/disturbed clusters with high purity (see their Table 1). 

We do not do exactly that, but we decided to adopt different cuts (to separate relaxed and disturbed systems) than those provided by \citet{and17} for the four X-ray parameters. Hence, we searched for the optimal values of those parameters (C$_{SB}$, C$_{SB4}$, $\delta$ and n$_{core}$). For each parameter we vary its value adopted to split the cluster population in relaxed and disturbed, building the distributions for the other three parameters. Then we chose the best separation of the two populations according to the KS test. The original cuts (or break values) adopted by \citet{and17} are C$_{SB}$ =  0.40, C$_{SB4}$ = 0.075, $\delta$ = 0.50 and n$_{core}$ = 1.5$\times$10$^{-2}$ cm$^{-3}$. The optimal break values we found are C$_{SB}$ =  0.26, C$_{SB4}$ = 0.055, $\delta$ = 0.46 and n$_{core}$ = 8$\times$10$^{-3}$ cm$^{-3}$. The fraction of relaxed clusters obtained with these optimal values for the X-ray classification are listed in the the last two lines of Table \ref{tab:relaxed_frac}.

Although our samples are much smaller than the original ones used by \citet{and17}, we verify a good agreement (first two lines of Table \ref{tab:relaxed_frac}) with the fraction of CC clusters they reported. The only property showing a slight disagreement is the Cuspiness ($\delta$) for the ESZ clusters. We corroborate their results regarding the larger fraction of CC clusters in the X-ray selected systems, although the difference between the ESZ and X-ray fractions is less pronounced for the C$_{SB}$ measure. As expected, the results with the optimal break values show larger fractions of relaxed clusters, especially for C$_{SB}$. According to the results presented by \citet{lov17} our new cuts privilege the selection of more relaxed clusters. However, it is important to stress this was not our goal, as we simply searched for the best break values to split clusters in two populations (of relaxed and disturbed objects). Despite the adoption of the optimal break values, when comparing the EsZ and X-ray samples we still detect a larger fraction of relaxed clusters for the latter.

In Table \ref{tab:relaxed_frac} we can also see that with the optical tests we still find the nominal fraction of relaxed clusters to be higher in the X-ray sample. However, the results differ by more than 10\% only for the Lee 3D test and for the classification done according to the BCG$-$X-ray offset. This is one of the main results from our work: the corroboration of different cluster populations present in the SZ and X-ray selected samples. 

\begin{table*}

\begin{minipage}{145mm}

\caption{Percentage of clusters classified as relaxed according to each test for the ESZ and X-ray samples. The last two lines show the results with optimal break values for the X-ray classification (see text in $\S$ \ref{frac_rel}).}
\label{tab:relaxed_frac}

\begin{tabular}{@{}lccccccccccc}

\hline

  & $\beta$ & $\Delta$ & L3D & AD & Offset & $\Delta$m$_{12}$ & C$_{SB}$ & C$_{SB4}$ & $\delta$ & n$_{core}$ & N$_{cls}$\\



\hline

ESZ & 63 ${\pm}~ 8$ & 48 ${\pm}~ 8$ & 53 ${\pm}~ 8$ & 53 ${\pm}~ 8$ & 48 ${\pm}~ 8$ & 38 ${\pm}~ 8$ & 25 ${\pm}~ 7$ & 38 ${\pm}~ 8$ & 50 ${\pm}~ 8$ & 38 ${\pm}~ 8$ & 40 \\
X-ray & 66 ${\pm}~ 6$ & 56 ${\pm}~ 6$ & 65 ${\pm}~ 6$ & 55 ${\pm}~ 6$ & 66 ${\pm}~ 6$ & 47 ${\pm}~ 6$ & 35 ${\pm}~ 6$ & 61 ${\pm}~ 6$ & 68 ${\pm}~ 6$ & 52 ${\pm}~ 6$ & 62 \\
ESZ &  &  &  &  &  &  & 48 ${\pm}~ 8$ & 53 ${\pm}~ 8$ & 53 ${\pm}~ 8$ & 50 ${\pm}~ 8$ & 40 \\
X-ray &  &  &  &  &  &  & 65 ${\pm}~ 6$ & 69 ${\pm}~ 6$ & 68 ${\pm}~ 6$ & 69 ${\pm}~ 6$ & 62 \\

\hline

\end{tabular}


\end{minipage}

\end{table*}

We can also verify that the fractions of relaxed clusters derived from the optical tests are generally larger than the ones indicated by the X-ray properties, when considering the original break values. However, both the optical and the X-ray tests show a large variation. For instance, for the X-ray sample, the fraction of relaxed clusters varies from 35$\%$ to 68$\%$, while it goes from 25$\%$ to 50$\%$ for the ESZ sample. For the optical indicators the corresponding fractions of relaxed systems range from 47$\%$ to 66$\%$, for the X-ray sample, and 38$\%$ to 63$\%$ for the ESZ sample. If we consider the X-ray results obtained with the optimal break values we find fractions of relaxed clusters in better agreement with the optical classifications.


\subsection{Comparison of Substructure Measurements} \label{comp_sub}

In Table \ref{tab:sub_comparison} we show the comparison between all the substructure indicators. Note the X-ray classification considers the original break values of \citet{and17}. The fraction of agreement between two estimates for classifying clusters as relaxed or not relaxed is shown. The recovery rate of each of the four optical tests ($\beta$, $\Delta$, Lee 3D and AD) when compared to the four X-ray measures is about $\sim 55\%$, with slightly better results for the $\Delta$ (DS) test and worst for the AD. However, the best performances are given by the indicators related to the BCGs (Offset to the X-ray position and $\Delta m_{12}$). These indicators of the dynamical state of the cluster have a recovery rate of the X-ray estimates of $\sim 70\%$. This is another main result from this work: the BCG to X-ray offset and $\Delta m_{12}$ trace very well the dynamical state of clusters as indicated by the X-ray morphological parameters. As the two parameters are observationally cheap, especially $\Delta m_{12}$, this represents a straightforward way to classify clusters. It requires only the X-ray centroid and the BCG position, or the magnitudes of the two BCGs.

Table \ref{tab:sub_comparison2} is analogous to Table \ref{tab:sub_comparison}, but considering the X-ray classification with the optimal break values we discussed above. Hence, the comparison between optical parameters is not changed. We can see the recovery rate of each of the four optical tests ($\beta$, $\Delta$ and Lee 3D, AD) when compared to the four X-ray measures is slightly better now, being about $\sim 60\%$, the exception being the AD measure. Although the AD results are now improved, they still give a fraction of agreement of $\sim 50\%$ with the X-ray classifications. The best performances are again obtained by the BCG to X-ray Offset and $\Delta m_{12}$. In particular, the first one is much improved, giving a fraction of agreement of $\sim 75\%$. It is important to stress that the optimal break values lead to a better agreement among the X-ray classifications themselves. Before (Table \ref{tab:sub_comparison}), the recovery rate of each of the four X-ray indicators was $\sim 80\%$, and it is now $\sim 90\%$ (Table \ref{tab:sub_comparison2}).

\begin{table*}

\begin{minipage}{112mm}

\caption{Comparison between all the substructure estimates for the full sample with 72 clusters. Fraction of agreement for each pair of substructure (or CC) measures. Results based on the original X-ray classification (see text in $\S$ \ref{frac_rel}).}
\label{tab:sub_comparison}

\begin{tabular}{@{}lcccccccccc}

\hline

  & $\beta$ & $\Delta$ & L3D & AD & Offset & $\Delta$m$_{12}$ & C$_{SB}$ & C$_{SB4}$ & $\delta$ & n$_{core}$ \\



\hline

$\beta$ & 1.00 & 0.69 & 0.72 & 0.59 & 0.54 & 0.49 & 0.51 & 0.56 & 0.58 & 0.52 \\
$\Delta$ &      & 1.00 & 0.55 & 0.62 & 0.56 & 0.58 & 0.62 & 0.51 & 0.52 & 0.58 \\
L3D &      &      & 1.00 & 0.54 & 0.56 & 0.49 & 0.51 & 0.56 & 0.55 & 0.49 \\
AD &      &      &      & 1.00 & 0.52 & 0.51 & 0.44 & 0.44 & 0.51 & 0.51 \\
Offset &      &      &      &      & 1.00 & 0.68 & 0.63 & 0.72 & 0.73 & 0.65 \\
$\Delta$m$_{12}$ &      &      &      &      &      & 1.00 & 0.68 & 0.76 & 0.75 & 0.75 \\
C$_{SB}$ &      &      &      &      &      &      & 1.00 & 0.75 & 0.65 & 0.82 \\
C$_{SB4}$ &      &      &      &      &      &      &      & 1.00 & 0.90 & 0.85 \\
$\delta$ &      &      &      &      &      &      &      &      & 1.00 & 0.80 \\
n$_{core}$ &     &      &      &      &      &      &      &      &      & 1.00 \\

\hline

\end{tabular}


\end{minipage}

\end{table*}

\begin{table*}

\begin{minipage}{112mm}

\caption{Comparison between all the substructure estimates for the full sample with 72 clusters. Fraction of agreement for each pair of substructure (or CC) measures. Results based on the break optimal values for the X-ray classification (see text in $\S$ \ref{frac_rel}).}
\label{tab:sub_comparison2}

\begin{tabular}{@{}lcccccccccc}

\hline

  & $\beta$ & $\Delta$ & L3D & AD & Offset & $\Delta$m$_{12}$ & C$_{SB}$ & C$_{SB4}$ & $\delta$ & n$_{core}$ \\



\hline

$\beta$ & 1.00 & 0.69 & 0.72 & 0.59 & 0.54 & 0.49 & 0.55 & 0.59 & 0.56 & 0.58 \\
$\Delta$ &      & 1.00 & 0.55 & 0.62 & 0.56 & 0.58 & 0.61 & 0.54 & 0.51 & 0.52 \\
L3D &      &      & 1.00 & 0.54 & 0.56 & 0.49 & 0.58 & 0.62 & 0.56 & 0.61 \\
AD &      &      &      & 1.00 & 0.52 & 0.51 & 0.48 & 0.52 & 0.49 & 0.51 \\
Offset &      &      &      &      & 1.00 & 0.68 & 0.68 & 0.77 & 0.75 & 0.76 \\
$\Delta$m$_{12}$ &      &      &      &      &      & 1.00 & 0.72 & 0.70 & 0.73 & 0.72 \\
C$_{SB}$ &      &      &      &      &      &      & 1.00 & 0.85 & 0.85 & 0.89 \\
C$_{SB4}$ &      &      &      &      &      &      &      & 1.00 & 0.94 & 0.96 \\
$\delta$ &      &      &      &      &      &      &      &      & 1.00 & 0.96 \\
n$_{core}$ &      &      &      &      &      &      &      &      &      & 1.00 \\

\hline

\end{tabular}


\end{minipage}

\end{table*}

As mentioned in $\S$ \ref{bcg_off} we adopt a different criterion than \citet{lav16} to classify clusters (as relaxed or not) based on the BCG$-$X-ray offset. In a similar way to what was described for the optimal choice of the X-ray break values, we allowed the BCG$-$X-ray offset break value to vary while splitting the cluster population in relaxed and disturbed, building the distributions for the four X-ray parameters. Next we selected the best separation of the two populations according to the KS test. We found that a cut at 0.01$\times$R$_{500}$ leads to a much better separation than using the 0.05$\times$R$_{500}$ value listed in \citet{lav16}. 
Actually the best results were found for a slightly smaller offset (0.008$\times$R$_{500}$). The new choice naturally reflects in the recovery rate of the X-ray results. Using the former value (0.05$\times$R$_{500}$) the percentage of agreement between the BCG$-$X-ray offset classification and the four X-ray properties is 47$\%$ (C$_{SB}$), 64$\%$ (C$_{SB4})$, 71$\%$ ($\delta$) and 54$\%$ (n$_{core}$). As we can see from Table \ref{tab:sub_comparison} those numbers are, for the offset cut proposed in this paper, 63$\%$ (C$_{SB}$), 72$\%$ (C$_{SB4}$), 73$\%$ ($\delta$) and 65$\%$ (n$_{core}$), being even better in Table \ref{tab:sub_comparison2}. A similar search for the best break value of $\Delta m_{12}$ leads to the conclusion that we can reliably split relaxed and disturbed clusters, according to the magnitude gap, at $\Delta m_{12} = 1.0$. Relaxed clusters have $\Delta m_{12} > 1.0$, while the opposite is true for disturbed systems.

\begin{figure}
\begin{center}
\leavevmode
\includegraphics[width=3.5in]{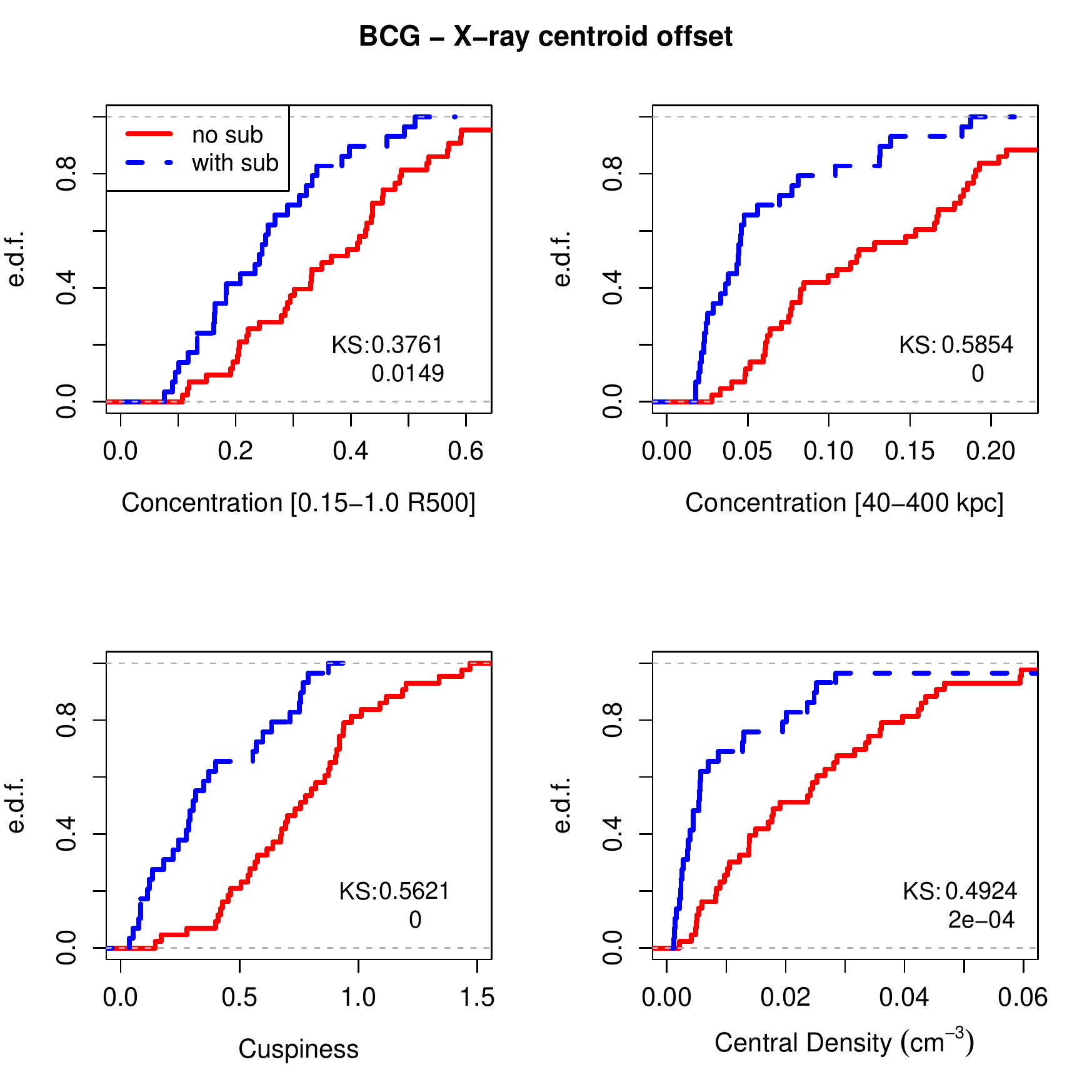}
\end{center}
\caption{Distribution of the four X-ray structural parameters. On the top left we show the concentration within $0.15-1.0$R$_{500}$, while in the top right the results are for the concentration within $40-400$kpc. In the bottom left we have the cuspiness distribution, while the central density is on the bottom right. On all panels clusters are classified as relaxed (solid red lines) or with substructure (blue dashed lines) according to the BCG$-$X-ray centroid offset. The KS and $p-$values are shown on all panels.}
\label{fig:edf_Xpars_offbcg}
\end{figure}

In Fig.~\ref{fig:edf_Xpars_offbcg} we show the distribution of the four X-ray parameters. Relaxed clusters are displayed by solid red lines, while disturbed systems are in blue dashed lines. The separation is done according to the BCG$-$X-ray centroid offset. The KS and $p-$values are indicated on all panels. It is well known that the BCG$-$X-ray offset provides a very reliable way to classify the dynamical state of clusters \citep{jon84,pos95,lin04,lav16}. This is verified in Fig.~\ref{fig:edf_Xpars_offbcg}, but we further show that the threshold adopted should be very small (0.01$\times$R$_{500}$), as this value results in a much better separation of clusters according to all the four X-ray properties.

\begin{figure}
\begin{center}
\leavevmode
\includegraphics[width=3.5in]{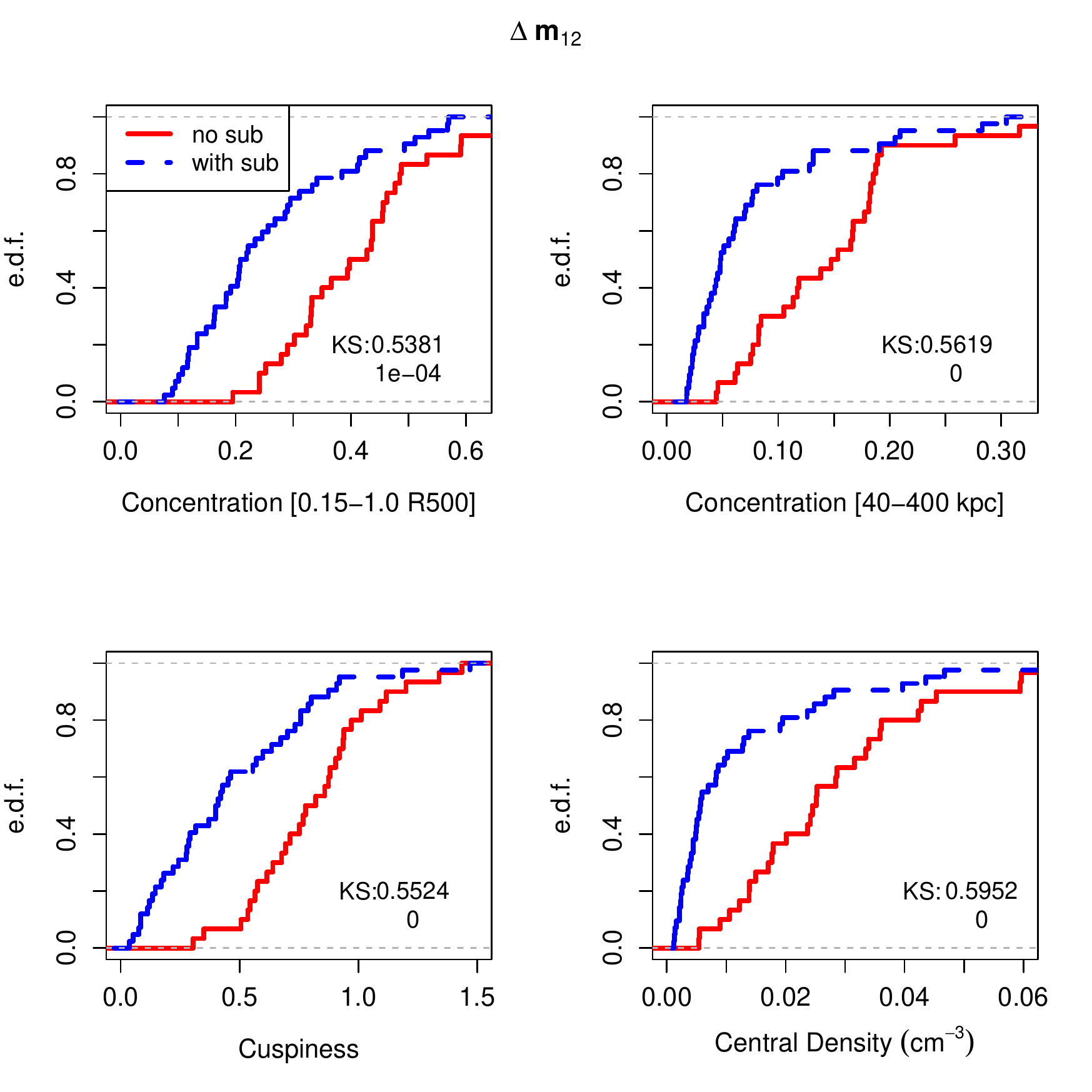}
\end{center}
\caption{Analogous to the previous figure, except for the way clusters are classified as relaxed or not. Instead of using the BCG$-$X-ray offset we consider the magnitude gap between the first and second brightest galaxies.}
\label{fig:edf_Xpars_mag12}
\end{figure}

In Fig.~\ref{fig:edf_Xpars_mag12} we show similar results to Fig.~\ref{fig:edf_Xpars_offbcg}, but the cluster separation into relaxed or disturbed systems is done according to the magnitude gap between the first and second brightest cluster galaxies. We see that this parameter is also a reliable proxy to the cluster dynamical state, like the BCG$-$X-ray offset described above. The main advantage of this method is that it depends only on the proper selection of the two brightest galaxies within the cluster. It does not even require the availability of X-ray data. It is not observationally expensive like optical substructure tests, because it does not require redshifts of a large number of galaxies in the cluster. The magnitude gap is, therefore, a cheap and reliable classification method for situations where a simple assessment of the cluster evolutionary state, and not the global cluster structure, is desired. If the X-ray centroid is also known the BCG$-$X-ray offset can also be employed.

Our main results regarding the comparison of morphological estimators seem to be in disagreement with the recent work of \citet{rob18}, who also compared X-ray and optical parameters to assess the cluster dynamical state. They find, for instance, that the offset between the BCG and cluster centroid is not a good tracer of cluster relaxation, but the AD test is. This is nearly opposite to our conclusion. Unfortunately, the only identical parameter between our work and theirs is the AD statistic, making a direct comparison of two pairs of parameters impossible. Besides AD, they used two other optical relaxation proxies: the ``stellar mass ratio'' (M$_2$/M$_1$) between the two brightest cluster galaxies, and the projected
offset between the BCG and the luminosity-weighted cluster centre ($\delta R_{MMG}$). In X-rays they use the Photon asymmetry (A$_{phot}$) and the centroid shift ($w$). In principle, M$_2$/M$_1$ should be similar to $\Delta$m$_{12}$, but it is subject to larger errors from the SED fitting procedure. The same can be said for $\delta R_{MMG}$, that should correlate well with the BCG X-ray centroid offset. However, the choice of a luminosity-weighted cluster centre (instead of the X-ray centroid) may not be the best, as the luminosity-weighted measure may be affected by issues like incompleteness in the spectroscopic data ({\it e.g.} fiber collision issue).

However, perhaps the most important difference from our work is the cluster sample. \citet{rob18} use the Yang  catalog, with at least 10 members, resulting in a sample with most objects more massive than $10^{14}$ M$_{\odot}$, up to $z = 0.2$. From Fig.~\ref{fig:boxplot} below we can see the Yang catalog at M $> 10^{14}$ M$_{\odot}$ is still much less massive than the ESZ and X-ray samples we use. More important is the fact that the SDSS main redshift survey is complete to $z = 0.1$. At this redshift the survey is complete to M$^*+1$. As pointed out in \citet{lop09a} a proper assessment of the cluster mass requires completeness at least to M$^*+1$. Hence, at $z > 0.1$, even having a large number of galaxies, the velocity dispersion, radius and mass estimates may be biased. The same is probably true for the substructure estimates. Although they claim no difference in their results for clusters at $z > 0.1$ or $z < 0.1$ we think the use of SDSS $z > 0.1$, and all the differences in the tests and samples mentioned above, prevent a proper comparison to our work.

\subsection{Central Galaxy Paradigm} \label{CGP}

A common assumption made in many studies is the central galaxy paradigm (CGP). According to it, in a dark matter halo the brightest (and most massive) halo galaxy (BHG) resides at rest at the centre of the dark matter potential well.

Many different studies investigated if the CGP is valid. Although they find most BCGs are nearly at rest at the central position of the parent cluster, some of them are not \citep{pos95, lin04, coz09}. In particular, \citet{van05} and \citet{ski11} find evidence for violation of the central galaxy paradigm, with the latter pointing to its validity only for very low mass haloes (M $\lesssim 10^{13}$ h$^{-1}$ M$_{\odot}$). A detailed discussion of this topic is beyond the goal of this paper. However, we still investigate if the velocity offsets of the BCGs are small (consistent with the CGP) or not. 

\begin{figure}
\begin{center}
\leavevmode
\includegraphics[width=3.5in]{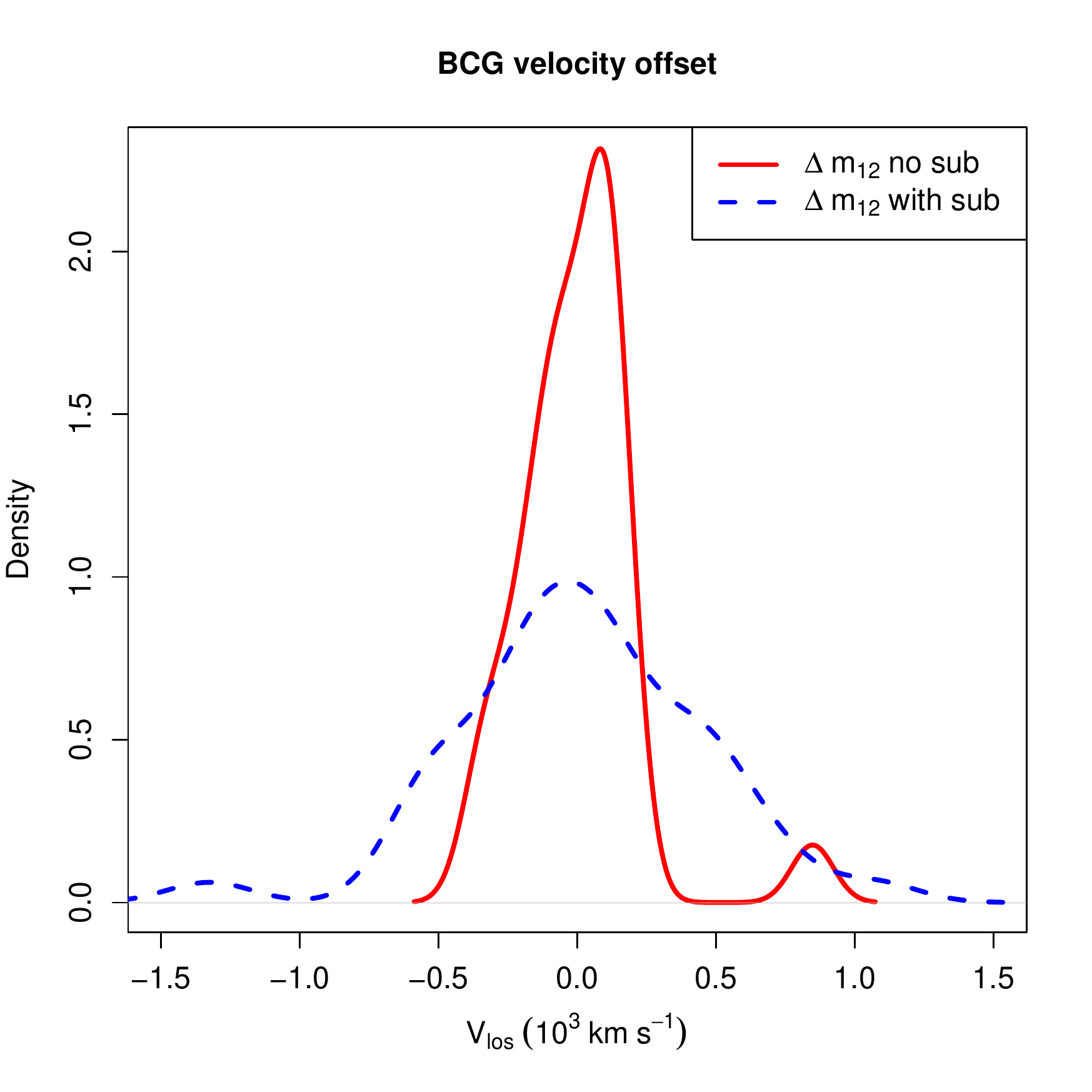}
\end{center}
\caption{BCG velocity offset distributions for relaxed (solid red lines) and disturbed (blue dashed lines) systems according to the magnitude gap classification.}
\label{fig:voff_mag12}
\end{figure}

In Fig.~\ref{fig:voff_mag12} we show the BCG velocity offsets relative to their parent clusters. We do so for relaxed (solid red lines) and disturbed (blue dashed lines) clusters, using the classification from the magnitude gap ($\Delta$m$_{12}$). We see that for relaxed clusters the BCG have small velocity offsets most of times, being consistent with the CGP. For the disturbed clusters the distribution is broader, so that a large fraction of BCGs are not at rest relative to their parent systems. In particular, from all 72 clusters in our sample we have 14\% with BCG absolute velocity offsets $>$ 500 km $s^{-1}$ and 42\% with $|\Delta_v| >$ 200 km s$^{-1}$. The latter fraction is also found using a cut in normalized velocity offset (|$\Delta_v/\sigma_v| >$ 0.30). If we consider only the relaxed clusters (according to the $\Delta$m$_{12}$ classification)
we have 20\% of clusters (BCGs) with $|\Delta_v| >$ 200 km s$^{-1}$ (or $|\Delta_v/\sigma_v| >$ 0.30). For the
disturbed clusters we find 57\% objects with the same velocity differences.

Using a large sample of 452 BCG dominated Abell clusters, \citet{coz09} found that the BCGs have a median normalized velocity offset of 0.32. In our sample, dominated by rich massive clusters, we find a median value $|\Delta_v/\sigma_v| =$ 0.23. For the relaxed systems this value decreases to 0.18, while it is 0.34 for the disturbed systems. \citet{coz09} conclude the BCG peculiar velocity is not dependent of the cluster richness and slightly depends on the Bautz$-$Morgan type. It is not our goal to make a detailed comparison with their results. However, from our results above, it seems their sample is largely dominated by disturbed objects, as the median $|\Delta_v/\sigma_v|$ for our full sample is smaller than theirs, but agrees with them for our subset of disturbed systems.

We therefore conclude that for massive systems the CGP may not be valid for a considerable fraction of objects, depending on the maximum velocity separation you allow. For $|\Delta_v/\sigma_v| >$ 0.30 this fraction is 42\%, raising to 57\% for the disturbed clusters and being 20\% for the relaxed systems. Although 1/5 
could be thought as a large number of relaxed clusters violating the CGP it is important to keep in mind some residual oscillation (due to previous major mergers) may persist long after the overall cluster relaxation \citep{har17}.

\section{Discussion}


With the availability of the cluster samples recently selected by the Planck satellite, it has been realized that they present fewer CC or relaxed systems in comparison to flux limited X-ray samples \citep{plc11,ros17}. The different dependence of the X-ray emission and SZ signal on the gas density could be responsible for the discrepant results. The X-ray emission scales with the square of the gas density, while SZ surveys are less sensitive to the central gas density. Hence, X-ray surveys are more likely to select more centrally peaked, relaxed systems (known as CC bias, \citealt{eck11}) at fixed mass, in comparison with SZ experiments. Additionally, this Malmquist bias also leads to an increased scatter of the mass-luminosity relation of flux limited X-ray samples \citep{cho12}.

Nonetheless, recent results from \citet{cho17} indicate that the main reason for the different fractions of CC clusters in X-ray and SZ samples is the fact that the X-ray samples are constructed from flux limited surveys, in opposition to a mass limited, nearly distance independent, SZ selection. They reach that conclusion by deriving morphological parameters and CC fractions of clusters in an X-ray volume limited sample, as well as two X-ray flux limited samples.

In this work we used an X-ray flux limited and an ESZ samples. However, both were limited in redshift ($z = 0.11$) due to the incompleteness in galaxy samples with spectroscopic redshifts above this limit. Even with this constraint we still find larger fractions of CC clusters in the X-ray selected sample when using the X-ray CC parameters from \citet{and17}. The results based on the optical substructure estimates tend to agree with that, except for the $\beta$ and AD tests. The Lee 3D and the BCG$-$X-ray offset show the largest differences between the ESZ and X-ray relaxed fractions.


At first, it seems the results from this work (using optical substructure estimates) are discrepant when compared to what is inferred from the X-ray measurements. 
However, as seen in Table \ref{tab:relaxed_frac} there is a large variation in the fractions of CC obtained from the four X-ray properties for the two samples. The difference between optical and X-ray results is also smaller when considering the optimal break values (last two lines of Table \ref{tab:relaxed_frac}). From Tables \ref{tab:sub_comparison} and \ref{tab:sub_comparison2} we also see the fraction of agreement of the X-ray measures could show a large variation, from 65 to 90\% (with the original break values considered in Table \ref{tab:sub_comparison}).

\begin{figure}
\begin{center}
\leavevmode
\includegraphics[width=3.5in]{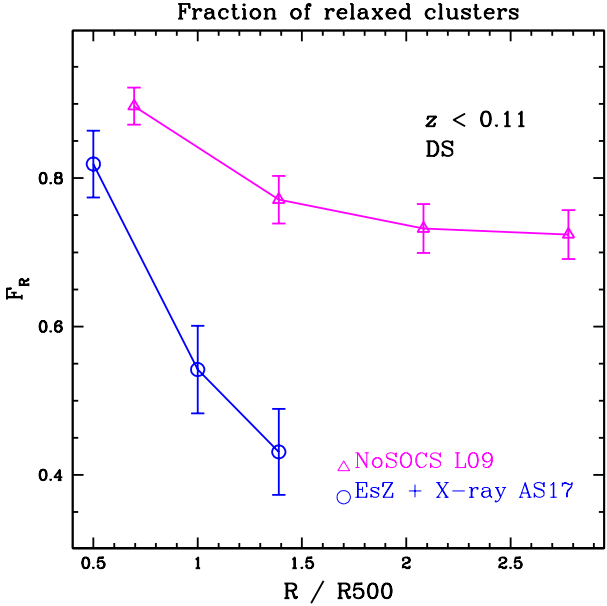}
\end{center}
\caption{Fraction of relaxed clusters as a function of the radius used for estimating substructure with the DS test. We show the results for the combined ESZ plus X-ray samples at $z < 0.11$ (circles), as well as the results for the NoSOCS sample (triangles) with 183 clusters described in \citet{lop09a}. The results for the current work (ESZ$+$X-ray) were originally obtained within R$_{500}$ and here we also show within 0.5$\times$R$_{500}$ and R$_{200}$. The NoSOCS sample had substructure measurements within R$_{200}$ and we also show here with 0.5, 1.5 and 2.0$\times$R$_{200}$. For our data we find R$_{200} \sim 1.39\times$R$_{500}$. We use that scale to display the above results as a function of R$_{500}$.}
\label{fig:fsub_vs_rad}
\end{figure}

Nonetheless, we tried to minimize possible differences in the optical and X-ray substructure estimates. For instance, we ran the four optical tests ($\beta$, $\Delta$ and Lee 3D, AD) within R$_{500}$, the same aperture used for the X-ray measurements. It is well known that larger apertures could result in smaller fractions of relaxed clusters as we probe non-equilibrium regions, dominated by infalling populations. This can be seen in Fig.~\ref{fig:fsub_vs_rad}, where we display, as an example, the
fraction of relaxed clusters as a function of the aperture adopted to measure substructure with the $\Delta$ test. The NoSOCS sample shows a variation of relaxed systems of $\sim$ 90 \% to 72 \%, from $\sim$ 0.5 to 3.0$\times$R$_{500}$. The ESZ$+$X-ray samples vary from $\sim$82\% to 43\%, between $\sim$ 0.5 and 1.5$\times$R$_{500}$. From this plot we can see that the fraction of relaxed clusters of the two samples are very different, being smaller for the ESZ$+$X-ray sample. The radial dependence is also more pronounced for this data set. 

\begin{figure}
\begin{center}
\leavevmode
\includegraphics[width=3.5in]{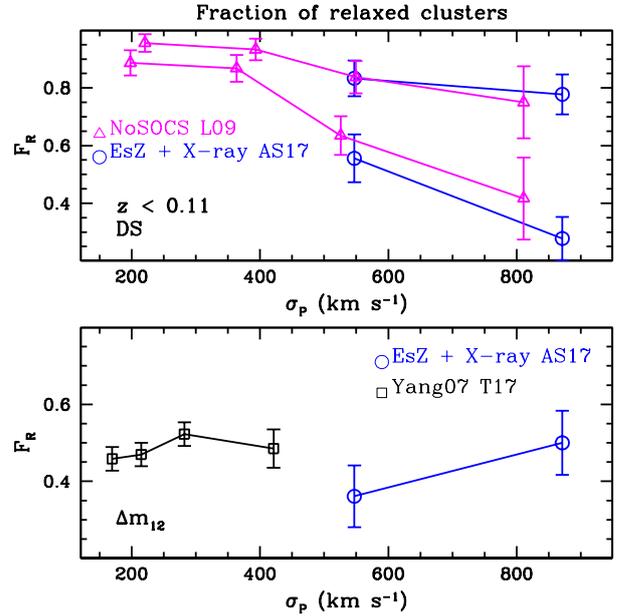}
\end{center}
\caption{Fraction of relaxed clusters as a function of cluster velocity dispersion. In the top panel we show the $\Delta$ test results for the NoSOCS (triangles) sample \citep{lop09a} and the combined ESZ plus X-ray samples (circles) at $z < 0.11$. For each sample we show two results, computed at two different apertures. For both samples the lower fractions were obtained within R$_{200}$. The higher fractions were derived using 0.5$\times$R$_{200}$ (NoSOCS) and 0.5$\times$R$_{500}$ (ESZ$+$X-ray). In the bottom panel we display the results for the magnitude gap for the ESZ$+$X-ray sample (circles), as well as the sample described in \citet{tre17b} (from which we used the 893 groups with $\sigma_P \ge 150$ $km$ $s^{-1}$). This sample is displayed with the squares.}
\label{fig:fsub_vs_vdisp}
\end{figure}

From this result we infer that the mass limit of the different samples can probably also explain the discrepant fractions of relaxed objects, at least on what regards the results based on the DS test. This is further explored in Fig.~\ref{fig:fsub_vs_vdisp}, where we show the variation of the fraction of relaxed clusters as a function of velocity dispersion, for the DS test (top) and magnitude gap (bottom). The top panel displays the results for the NoSOCS and ESZ$+$X-ray samples, in each case obtained with two apertures. For both samples the lower fractions were computed within R$_{200}$, while the higher fractions were derived with 0.5$\times$R$_{200}$ (NoSOCS) and 0.5$\times$R$_{500}$ (ESZ$+$X-ray). In the bottom panel we show the results for the ESZ$+$X-ray plus the sample of \citet{yan07}. From the \citet{yan07} catalog we only use objects with $\sigma_P \ge 150$ km s$^{-1}$. The first and second BCGs were selected as in \citet{tre17b}, but using the same aperture as in the current work. We can see the $\Delta$ test results vary with mass (velocity dispersion), while those from the magnitude gap are nearly constant with $\sigma_P$. This last measurement is central, while the first depends on the galaxy distribution up to R$_{200}$, hence, being more sensitive to infalling populations. 

From the upper panel of Fig.~\ref{fig:fsub_vs_vdisp} we can also verify some other interesting points. First, the variation with $\sigma_P$ is significant only for systems with $\sigma_P \gtrsim 400$ km s$^{-1}$. Second, the mass dependence of the fraction of relaxed clusters is much less pronounced within a small aperture. So, if the substructure test is restricted to the central region of clusters (which is generally the case for X-ray data) it may be impossible to detect a variation with mass, especially if the analysis is based on massive systems, such as in the ESZ$+$X-ray sample (blue circles). Finally, it is important to mention there is no bias in the mass correlation we show, as we adopted a large aperture (R$_{200}$) that scales with mass. However, even the results within R$_{500}$ are still consistent with a mass dependence of the fraction of relaxed clusters.

We have also found some indication that the C$_{SB}$ and C$_{SB4}$ parameters depend on cluster mass, while $\delta$ and n$_{core}$ do not. However, due to the small sample size the error bars are large. Nonetheless, it is important to mention that $\delta$ and n$_{core}$ are central measurements, which are not sensitive to large scale variations in the gas distribution. In this case, the size of the cluster is not important. Similar conclusions are also reached for the $\beta$ test and the BCG$-$X-ray offset.



\begin{figure}
\begin{center}
\leavevmode
\includegraphics[width=3.5in]{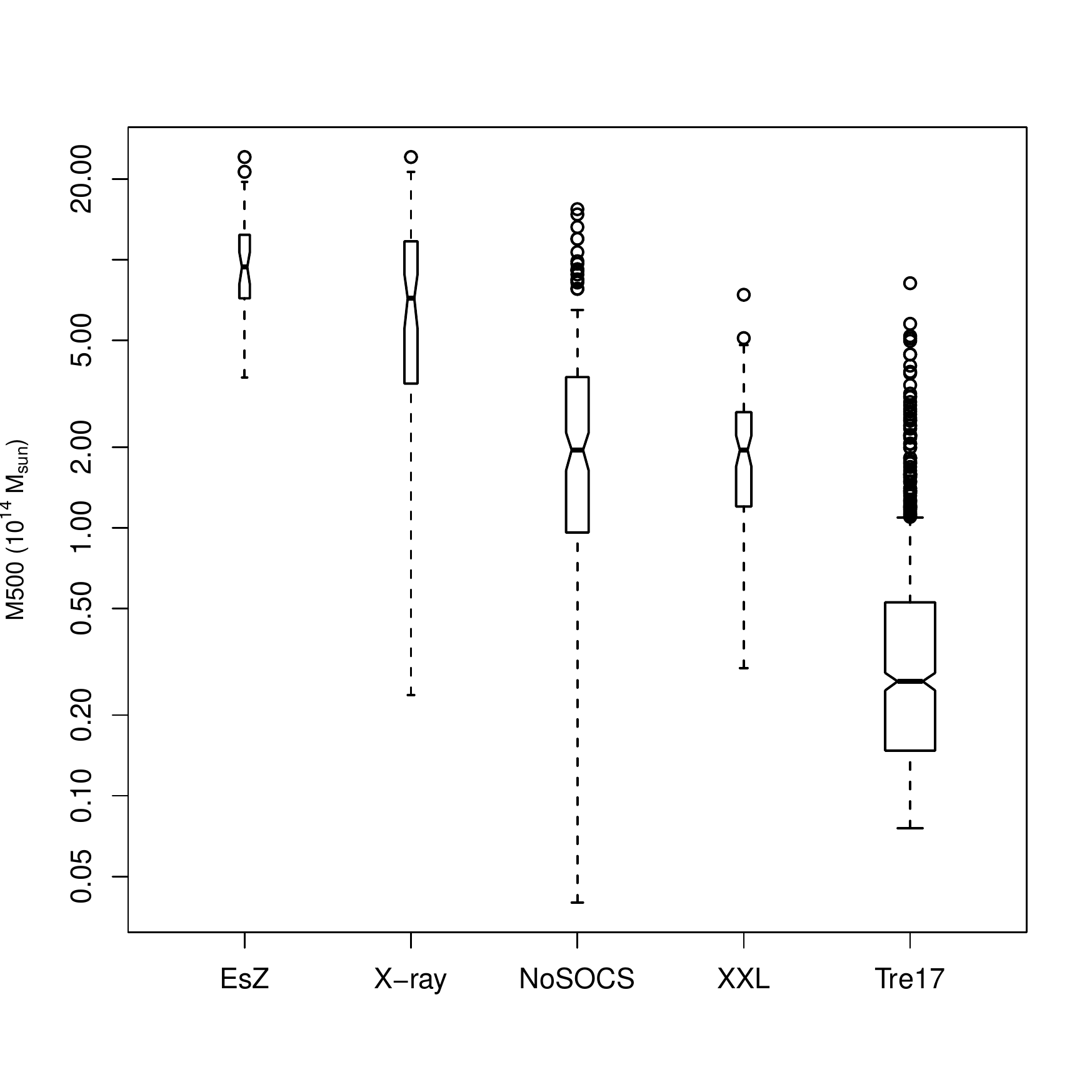}
\end{center}
\caption{Mass boxplot of the samples seen in Figs. \ref{fig:fsub_vs_vdisp} and \ref{fig:fsub_vs_z}. Data points outside the whiskers are plotted individually as outliers. The box width scales with the square root of each sample size.}
\label{fig:boxplot}
\end{figure}

\begin{figure}
\begin{center}
\leavevmode
\includegraphics[width=3.5in]{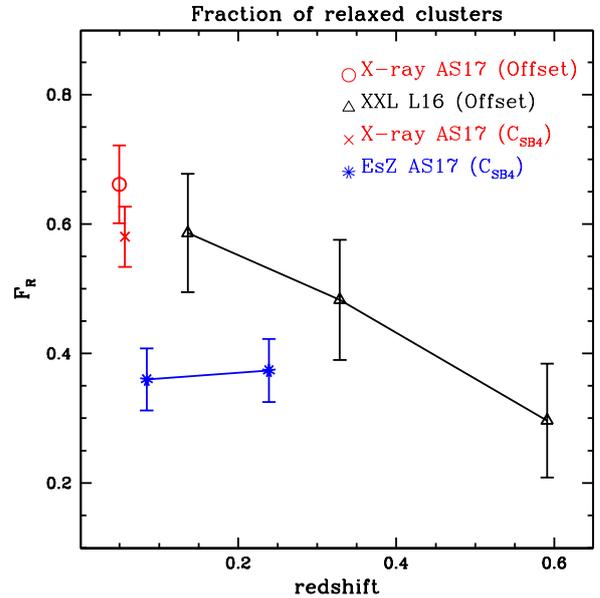}
\end{center}
\caption{Fraction of relaxed clusters as a function of redshift. We show the results for the 62 lower redshift X-ray clusters analysed in the current work with an open circle. Results from the XXL sample (85 clusters, \citealp{lav16}) are displayed with the open triangles. These two data sets consider the BCG$-$X-ray offset. For the XMM data we separate relaxed from disturbed clusters at 0.03$\times$R$_{500}$. The cross shows the result for the full X-ray sample (112 objects) and the stars indicate the ESZ (199 cluster) data from \citet{and17}. These two results are based on the C$_{SB4}$ parameter.}
\label{fig:fsub_vs_z}
\end{figure}

In Fig.~\ref{fig:boxplot} we show the cluster mass boxplot of the samples displayed in Figs. \ref{fig:fsub_vs_vdisp} and \ref{fig:fsub_vs_z}. We can see that the ESZ and X-ray samples of \citet{and17} are the most massive, with the first having the largest median mass value and largest fraction of massive objects. The NoSOCS and XXL ({\it Extragalactic observation program of the space mission XMM Newton}) samples have similar median masses, but the former is more extended to low and high mass regimes. The group sample from \citet{yan07} is by far the least massive one, not extending further to the lower masses as we cut the sample at $\sigma_P = 150$ $km$ $s^{-1}$.

Another possible explanation for the larger fraction of disturbed systems in the SZ sample in \citet{and17} is the higher redshift limit ($z \sim 0.3$) of that sample compared to the X-ray one ($z \sim 0.1$). In a hierarchical Universe it is expected that the fraction of clusters with substructure, as well as cluster morphology \citep{ho06}, evolve with redshift. As the Universe ages, the fraction of relaxed systems should increase, with most disturbed systems being the more massive ones. This morphological evolution is seen in simulations \citep{ho06}, but some observational results indicate little or zero redshift evolution \citep{wei13,nur17}. As we have already shown the fraction of relaxed systems is smaller for massive objects (Figs.~\ref{fig:fsub_vs_rad} and \ref{fig:fsub_vs_vdisp}), we now investigate the possible substructure dependence with redshift. 

In Fig.~\ref{fig:fsub_vs_z} we display the redshift variation of three samples. The ESZ and X-ray ones from \citet{and17}, as well as the XXL \citep{lav16}. For the two samples from \citet{and17} we used the C$_{SB4}$ parameter. In that case we could consider the full data sets (199 objects for the ESZ and 112 in the X-ray). For those two data sets (but at $z < 0.11$) and for the XXL data we also considered the BCG$-$X-ray offset. Due to the larger angular error in the X-ray centroid from XMM, we decided to use an intermediate criterion between theirs \citep{lav16} and ours to separate relaxed and disturbed systems for XXL. The separation is done at 0.03$\times$R$_{500}$. 

We can see a steep evolution in the fraction of relaxed clusters with redshift in the XXL data (from $z \sim 0.6$ to $\sim 0.15$). The X-ray sample (using the BCG$-$X-ray offset) from \citet{and17} at $z \sim 0.05$ is consistent with the XXL results. Considering C$_{SB4}$ the ESZ sample shows a constant behaviour between $z \sim 0.24$ and $\sim 0.08$, while the X-ray sample \citep{and17} is still consistent with the evolution inferred from XXL. Hence, even though we see a redshift evolution down to $z \sim 0.1$ from the XXL survey, the ESZ sample shows no evolution in the redshift range common with the XXL data. This is due to the fact that the ESZ clusters represent a nearly mass limited sample, independent of redshift. So, its substructure fraction reflects the typical values of clusters of that mass range. In other words, we can not claim that the substructure fraction difference between ESZ and X-ray clusters from \citet{and17} is due to the different redshifts probed. The main reason for the substructure fraction difference is thus, probably, the mass.

\section{Conclusions}

In this work we investigated if the fraction of relaxed clusters in an X-ray selected sample is larger than in a SZ catalog. This result has been found before from X-ray estimates of the cluster dynamical state. For the first time we show that is also the case when considering optical properties to trace cluster substructure (Table \ref{tab:relaxed_frac}). 

The SZ and X-ray samples we used are from \citet{and17}. The first has clusters up to a larger redshift limit ($z \sim 0.35$) and is composed of more massive clusters (see Fig.~\ref{fig:boxplot}). In order to have enough redshifts per cluster we only studied objects at $z < 0.11$, in agreement with the completeness limits of the surveys we used (SDSS, 2dFGRS and 6dF). We also added redshifts from NED to our data. After eliminating interlopers and selecting cluster members we performed a virial analysis, deriving velocity dispersion, physical radius and mass for all objects. The final SZ sample has 40 clusters, while the X-ray one has 62, and the combined data set comprises 72 systems. We employed four optical substructure tests to our data ($\beta$, $\Delta$, Lee 3D and AD). We have also estimated the dynamical state of the clusters from the BCG$-$X-ray offset, and from the magnitude gap between the first and second BCGs.

Besides using the optical substructure estimates to corroborate the higher fraction of relaxed systems in X-ray selected samples compared to SZ ones, we investigate if cluster mass plays a role on that. We found that optical substructure tests applied to the whole galaxy distribution depend on cluster mass (even within R$_{500}$), but centrally located tests do not (Figs.~\ref{fig:fsub_vs_rad} and~\ref{fig:fsub_vs_vdisp}). In the first case we have the $\beta$, $\Delta$ and Lee 3D tests, while those based on the BCG lie in the second class. Similar results are found for the X-ray substructure estimates (C$_{SB}$, C$_{SB4}$, $\delta$ and n$_{core}$). We have also shown that cluster substructure evolves with redshift (Fig.~\ref{fig:fsub_vs_z}), but that does not explain the higher fraction of disturbed objects in the higher-z SZ sample compared to the X-ray one.

Finally, we obtained a good agreement ($\sim 60 \%$) between the optical and X-ray substructure estimates. Nonetheless, the agreement is better for the substructure inferred from the BCG, either using the BCG$-$X-ray offset, or the magnitude gap. Hence, we advocate using those estimates as the most reliable and cheap way to assess a cluster dynamical state. However, we noticed that the BCG$-$X-ray offset threshold should be smaller than normally used in the literature. We found an optimal cut at $\sim$0.01$\times$R$_{500}$. Regarding the magnitude gap we separate relaxed from disturbed clusters at $\Delta m_{12} = 1.0$\footnote{Note that the $\Delta m_{12}$ limit to separate relaxed and non-relaxed clusters may vary with the maximum distance allowed for obtaining the second brightest galaxy, which in this work is $R_{\rm max} = 0.5\, R_{500}$. For a discussion on how the magnitude gap varies with $R_{\rm max}$ see, e.g., \citet{tre17b}.}. We should also note the above values obtained in the current study are based on a small redshift range ($z < 0.11$). 

We plan to extend this analysis to higher redshifts. We also aim to perform a comparison of cluster mass estimates obtained in the optical (using galaxy velocities and weak lensing) and in X-rays. Another goal for a forthcoming work is to investigate the mass calibration of these different samples and the impact of substructure on the mass estimates \citep{lop09b}. Finally, we will investigate a possible correlation between the BCG properties (S\'ersic index, star formation rate, etc.) and the cluster dynamical state. The characterization of the dynamical state of clusters and its impact on the mass calibration are important steps for the proper use of clusters as cosmological probes from large scale surveys.

\section*{Acknowledgements}

PAAL thanks the support of CNPq, grant
308969/2014-6; and CAPES, {\it Programa Est\'agio S\^enior no Exterior},
process number 88881.120856/2016-01. ALBR thanks for the support of CNPq, grant 311932/2017-7. 
F.D. acknowledges long-term support from CNES. TFL acknowledges financial support from CNPq (grant 303278/2015-3).
M.T. thanks the support from the PSL Research University.

This research has  made use of the SAO/NASA  Astrophysics Data System, the NASA/IPAC Extragalactic  Database (NED) and the ESA Sky tool (sky.esa.int/).  Funding for the SDSS
and  SDSS-II was  provided  by  the Alfred  P.  Sloan Foundation,  the
Participating  Institutions,  the  National  Science  Foundation,  the
U.S.  Department  of  Energy,   the  National  Aeronautics  and  Space
Administration, the  Japanese Monbukagakusho, the  Max Planck Society,
and  the Higher  Education  Funding  Council for  England.  A list  of
participating  institutions can  be obtained  from the  SDSS  Web Site
http://www.sdss.org/. The 2dF Galaxy Redshift Survey (http://www.2dfgrs.net/) has been made possible by the dedicated efforts of the staff of the Anglo-Australian Observatory, both in creating the 2dF instrument and in supporting it on the telescope. We are also very thankful for the availability of the Final Release of 6dFGS (http://www-wfau.roe.ac.uk/6dFGS/)




\bibliographystyle{mnras}
\bibliography{biblio_op_sub_sz_xray.bib} 




\appendix

\section{Tables with cluster and BCG properties}

\begin{table*}

\begin{minipage}{135mm}

\caption{Main properties of the 72 clusters.}
\label{tab:cls_props}

{\scriptsize
\begin{tabular}{@{}l*8{c}}

\hline

 name & ra & dec & z$_{\odot}$ & $\sigma_P$ & R$_{500}$ & M$_{500}$ & R$_{200}$ & M$_{200}$ \\


 & (J2000) & (J2000) & & (km s$^{-1}$) & (Mpc) & (10$^{14}$M$_{\odot}$) & (Mpc) & (10$^{14}$M$_{\odot}$) \\

\hline

A2734 & 2.84036 & -28.85404 & 0.0613 & 532.2 $_{-23.2} ^{+30.4}$ & 0.78 $_{-0.02} ^{+0.03}$ & 1.43 $_{-0.13} ^{+0.16}$ & 1.07 $_{-0.03} ^{+0.04}$ & 1.48 $_{-0.13} ^{+0.17}$ \\
A85 & 10.45996 & -9.30265 & 0.0555 & 846.6 $_{-39.5} ^{+47.3}$ & 1.57 $_{-0.05} ^{+0.06}$ & 11.68 $_{-1.09} ^{+1.31}$ & 2.20 $_{-0.07} ^{+0.08}$ & 12.70 $_{-1.19} ^{+1.42}$ \\
A119 & 14.06754 & -1.24972 & 0.0441 & 750.0 $_{-28.6} ^{+34.0}$ & 1.33 $_{-0.12} ^{+0.25}$ & 6.96 $_{-1.91} ^{+3.94}$ & 1.84 $_{-0.17} ^{+0.35}$ & 7.43 $_{-2.04} ^{+4.21}$ \\
A193 & 21.28150 & 8.69999 & 0.0487 & 616.3 $_{-51.0} ^{+60.4}$ & 0.94 $_{-0.05} ^{+0.06}$ & 2.44 $_{-0.41} ^{+0.48}$ & 1.32 $_{-0.07} ^{+0.09}$ & 2.73 $_{-0.46} ^{+0.54}$ \\
EXO0422 & 66.46318 & -8.55954 & 0.0392 & 284.4 $_{-39.8} ^{+72.7}$ & 0.52 $_{-0.05} ^{+0.09}$ & 0.42 $_{-0.12} ^{+0.22}$ & 0.72 $_{-0.07} ^{+0.12}$ & 0.44 $_{-0.13} ^{+0.23}$ \\
A496 & 68.40797 & -13.26168 & 0.0326 & 686.7 $_{-25.8} ^{+30.9}$ & 1.24 $_{-0.11} ^{+0.23}$ & 5.64 $_{-1.52} ^{+3.06}$ & 1.72 $_{-0.15} ^{+0.31}$ & 5.99 $_{-1.61} ^{+3.25}$ \\
S0540 & 85.02782 & -40.83657 & 0.0365 & 477.2 $_{-100.1} ^{+189.7}$ & 0.89 $_{-0.13} ^{+0.24}$ & 2.06 $_{-0.87} ^{+1.64}$ & 1.23 $_{-0.17} ^{+0.33}$ & 2.20 $_{-0.93} ^{+1.75}$ \\
A548e & 87.15963 & -25.47783 & 0.0403 & 642.4 $_{-16.6} ^{+19.7}$ & 1.24 $_{-0.02} ^{+0.03}$ & 5.59 $_{-0.29} ^{+0.34}$ & 1.71 $_{-0.03} ^{+0.04}$ & 5.92 $_{-0.31} ^{+0.36}$ \\
A3376 & 90.42327 & -39.98886 & 0.0461 & 661.7 $_{-31.5} ^{+37.1}$ & 1.12 $_{-0.04} ^{+0.04}$ & 4.21 $_{-0.40} ^{+0.47}$ & 1.55 $_{-0.05} ^{+0.06}$ & 4.44 $_{-0.42} ^{+0.50}$ \\
bA3376 & 90.54907 & -39.95732 & 0.0462 & 655.8 $_{-28.9} ^{+37.1}$ & 1.12 $_{-0.03} ^{+0.04}$ & 4.20 $_{-0.37} ^{+0.48}$ & 1.55 $_{-0.05} ^{+0.06}$ & 4.42 $_{-0.39} ^{+0.50}$ \\
A3391 & 96.58534 & -53.69329 & 0.0554 & 952.0 $_{-34.1} ^{+40.9}$ & 1.63 $_{-0.04} ^{+0.05}$ & 12.89 $_{-0.93} ^{+1.11}$ & 2.27 $_{-0.05} ^{+0.07}$ & 14.05 $_{-1.01} ^{+1.21}$ \\
A3395 & 96.70078 & -54.54923 & 0.0514 & 867.2 $_{-24.1} ^{+37.5}$ & 1.46 $_{-0.03} ^{+0.04}$ & 9.31 $_{-0.52} ^{+0.81}$ & 2.03 $_{-0.04} ^{+0.06}$ & 10.02 $_{-0.56} ^{+0.87}$ \\
bA3395 & 96.90003 & -54.44629 & 0.0505 & 855.1 $_{-29.3} ^{+32.3}$ & 1.42 $_{-0.03} ^{+0.04}$ & 8.59 $_{-0.59} ^{+0.65}$ & 1.98 $_{-0.05} ^{+0.05}$ & 9.23 $_{-0.64} ^{+0.70}$ \\
bA754 & 137.19782 & -9.64124 & 0.0543 & 930.5 $_{-28.8} ^{+36.1}$ & 1.58 $_{-0.13} ^{+0.24}$ & 11.71 $_{-2.87} ^{+5.37}$ & 2.20 $_{-0.18} ^{+0.34}$ & 12.72 $_{-3.12} ^{+5.84}$ \\
A754 & 137.33525 & -9.68431 & 0.0543 & 903.5 $_{-30.8} ^{+36.6}$ & 1.53 $_{-0.13} ^{+0.23}$ & 10.68 $_{-2.64} ^{+4.90}$ & 2.13 $_{-0.18} ^{+0.33}$ & 11.56 $_{-2.86} ^{+5.31}$ \\
G269.51+26.42 & 159.17049 & -27.52653 & 0.0124 & 648.4 $_{-22.1} ^{+24.6}$ & 1.08 $_{-0.11} ^{+0.26}$ & 3.64 $_{-1.10} ^{+2.65}$ & 1.49 $_{-0.15} ^{+0.36}$ & 3.82 $_{-1.16} ^{+2.78}$ \\
G241.85+51.53 & 159.91688 & 5.16102 & 0.0695 & 640.2 $_{-31.3} ^{+47.3}$ & 1.25 $_{-0.04} ^{+0.06}$ & 5.91 $_{-0.58} ^{+0.88}$ & 1.73 $_{-0.06} ^{+0.09}$ & 6.28 $_{-0.62} ^{+0.93}$ \\
USGCS152 & 162.60879 & -12.84501 & 0.0154 & 179.0 $_{-24.7} ^{+40.8}$ & 0.44 $_{-0.04} ^{+0.07}$ & 0.24 $_{-0.07} ^{+0.11}$ & 0.60 $_{-0.06} ^{+0.09}$ & 0.24 $_{-0.07} ^{+0.11}$ \\
G172.88+65.32 & 167.91863 & 40.84118 & 0.0758 & 562.1 $_{-25.8} ^{+31.0}$ & 1.27 $_{-0.04} ^{+0.05}$ & 6.20 $_{-0.57} ^{+0.69}$ & 1.75 $_{-0.05} ^{+0.06}$ & 6.59 $_{-0.61} ^{+0.73}$ \\
bG243.57+67.76 & 173.21115 & 14.45586 & 0.0815 & 616.1 $_{-38.7} ^{+53.9}$ & 1.36 $_{-0.06} ^{+0.08}$ & 7.67 $_{-0.97} ^{+1.34}$ & 1.88 $_{-0.08} ^{+0.11}$ & 8.21 $_{-1.04} ^{+1.44}$ \\
G243.57+67.76 & 173.21315 & 14.49224 & 0.0818 & 629.2 $_{-41.1} ^{+53.0}$ & 1.37 $_{-0.06} ^{+0.08}$ & 7.85 $_{-1.03} ^{+1.33}$ & 1.90 $_{-0.08} ^{+0.11}$ & 8.41 $_{-1.10} ^{+1.42}$ \\
G234.59+73.01 & 176.18399 & 19.70444 & 0.0216 & 603.6 $_{-23.2} ^{+26.1}$ & 1.23 $_{-0.11} ^{+0.23}$ & 5.41 $_{-1.48} ^{+3.02}$ & 1.70 $_{-0.16} ^{+0.32}$ & 5.74 $_{-1.57} ^{+3.21}$ \\
ZwCl1215 & 184.42165 & 3.65612 & 0.0768 & 735.0 $_{-36.1} ^{+44.4}$ & 1.46 $_{-0.05} ^{+0.06}$ & 9.61 $_{-0.95} ^{+1.17}$ & 2.04 $_{-0.07} ^{+0.08}$ & 10.36 $_{-1.02} ^{+1.26}$ \\
bA3528s & 193.59224 & -29.01311 & 0.0539 & 963.9 $_{-43.9} ^{+52.3}$ & 1.67 $_{-0.05} ^{+0.06}$ & 13.92 $_{-1.27} ^{+1.51}$ & 2.33 $_{-0.07} ^{+0.08}$ & 15.22 $_{-1.39} ^{+1.66}$ \\
A3528n & 193.59252 & -29.01299 & 0.0539 & 963.9 $_{-43.9} ^{+52.3}$ & 1.67 $_{-0.05} ^{+0.06}$ & 13.92 $_{-1.27} ^{+1.51}$ & 2.33 $_{-0.07} ^{+0.08}$ & 15.22 $_{-1.39} ^{+1.66}$ \\
A3528s & 193.66960 & -29.22793 & 0.0545 & 896.7 $_{-38.4} ^{+45.2}$ & 1.62 $_{-0.05} ^{+0.05}$ & 12.65 $_{-1.09} ^{+1.28}$ & 2.26 $_{-0.06} ^{+0.08}$ & 13.78 $_{-1.19} ^{+1.39}$ \\
A1644 & 194.29821 & -17.40932 & 0.0471 & 924.0 $_{-41.5} ^{+44.6}$ & 1.48 $_{-0.04} ^{+0.05}$ & 9.64 $_{-0.87} ^{+0.93}$ & 2.06 $_{-0.06} ^{+0.07}$ & 10.40 $_{-0.94} ^{+1.01}$ \\
A3532 & 194.34180 & -30.36371 & 0.0558 & 640.9 $_{-25.1} ^{+33.9}$ & 1.33 $_{-0.03} ^{+0.05}$ & 7.12 $_{-0.56} ^{+0.75}$ & 1.85 $_{-0.05} ^{+0.07}$ & 7.60 $_{-0.60} ^{+0.80}$ \\
A1650 & 194.67264 & -1.76207 & 0.0842 & 731.2 $_{-32.0} ^{+38.6}$ & 1.45 $_{-0.04} ^{+0.05}$ & 9.47 $_{-0.83} ^{+1.00}$ & 2.02 $_{-0.06} ^{+0.07}$ & 10.20 $_{-0.89} ^{+1.08}$ \\
A1651 & 194.84308 & -4.19592 & 0.0848 & 836.4 $_{-30.1} ^{+33.2}$ & 1.64 $_{-0.04} ^{+0.04}$ & 13.61 $_{-0.98} ^{+1.08}$ & 2.29 $_{-0.06} ^{+0.06}$ & 14.84 $_{-1.07} ^{+1.18}$ \\
G057.33+88.01 & 194.94856 & 27.95189 & 0.0232 & 911.3 $_{-19.6} ^{+22.9}$ & 1.66 $_{-0.14} ^{+0.27}$ & 13.37 $_{-3.36} ^{+6.60}$ & 2.32 $_{-0.19} ^{+0.38}$ & 14.60 $_{-3.67} ^{+7.21}$ \\
A1736 & 201.75385 & -27.19644 & 0.0456 & 918.2 $_{-66.4} ^{+87.6}$ & 1.39 $_{-0.07} ^{+0.09}$ & 7.91 $_{-1.15} ^{+1.51}$ & 1.95 $_{-0.09} ^{+0.12}$ & 8.79 $_{-1.27} ^{+1.68}$ \\
A3558 & 201.98677 & -31.49551 & 0.0475 & 872.5 $_{-17.4} ^{+20.1}$ & 1.68 $_{-0.14} ^{+0.26}$ & 14.16 $_{-3.45} ^{+6.58}$ & 2.35 $_{-0.19} ^{+0.36}$ & 15.49 $_{-3.77} ^{+7.20}$ \\
bA3558 & 202.44904 & -31.60724 & 0.0490 & 890.3 $_{-17.4} ^{+21.4}$ & 1.70 $_{-0.14} ^{+0.25}$ & 14.59 $_{-3.48} ^{+6.54}$ & 2.38 $_{-0.19} ^{+0.35}$ & 15.98 $_{-3.82} ^{+7.16}$ \\
bA3562 & 202.86448 & -31.82171 & 0.0457 & 901.0 $_{-18.9} ^{+22.9}$ & 1.68 $_{-0.14} ^{+0.28}$ & 13.98 $_{-3.53} ^{+6.99}$ & 2.34 $_{-0.20} ^{+0.39}$ & 15.29 $_{-3.87} ^{+7.64}$ \\
A3560 & 203.11565 & -33.14266 & 0.0489 & 662.5 $_{-27.2} ^{+34.1}$ & 1.13 $_{-0.03} ^{+0.04}$ & 4.34 $_{-0.36} ^{+0.45}$ & 1.57 $_{-0.04} ^{+0.05}$ & 4.58 $_{-0.38} ^{+0.47}$ \\
A3562 & 203.39470 & -31.67291 & 0.0489 & 894.1 $_{-20.9} ^{+24.5}$ & 1.60 $_{-0.13} ^{+0.25}$ & 12.29 $_{-3.02} ^{+5.76}$ & 2.24 $_{-0.18} ^{+0.35}$ & 13.37 $_{-3.28} ^{+6.27}$ \\
A1767 & 204.02488 & 59.20234 & 0.0707 & 737.2 $_{-36.9} ^{+48.3}$ & 1.45 $_{-0.05} ^{+0.06}$ & 9.29 $_{-0.93} ^{+1.22}$ & 2.02 $_{-0.07} ^{+0.09}$ & 10.00 $_{-1.00} ^{+1.31}$ \\
A1775 & 205.45360 & 26.37219 & 0.0754 & 418.7 $_{-14.5} ^{+29.0}$ & 0.95 $_{-0.02} ^{+0.04}$ & 2.58 $_{-0.18} ^{+0.36}$ & 1.30 $_{-0.03} ^{+0.06}$ & 2.69 $_{-0.19} ^{+0.37}$ \\
A3571 & 206.86712 & -32.86611 & 0.0393 & 911.7 $_{-45.8} ^{+52.9}$ & 1.35 $_{-0.05} ^{+0.05}$ & 7.24 $_{-0.73} ^{+0.84}$ & 1.87 $_{-0.06} ^{+0.07}$ & 7.74 $_{-0.78} ^{+0.90}$ \\
A1795 & 207.21963 & 26.59200 & 0.0629 & 696.5 $_{-32.6} ^{+39.8}$ & 1.36 $_{-0.04} ^{+0.05}$ & 7.66 $_{-0.72} ^{+0.88}$ & 1.89 $_{-0.06} ^{+0.07}$ & 8.20 $_{-0.77} ^{+0.94}$ \\
A1831 & 209.81413 & 27.97623 & 0.0630 & 335.2 $_{-21.2} ^{+30.9}$ & 0.91 $_{-0.04} ^{+0.06}$ & 2.30 $_{-0.29} ^{+0.42}$ & 1.26 $_{-0.05} ^{+0.08}$ & 2.39 $_{-0.30} ^{+0.44}$ \\
bMKW8 & 219.59114 & 3.66998 & 0.0270 & 344.7 $_{-12.0} ^{+12.6}$ & 0.91 $_{-0.02} ^{+0.02}$ & 2.22 $_{-0.16} ^{+0.16}$ & 1.26 $_{-0.03} ^{+0.03}$ & 2.31 $_{-0.16} ^{+0.17}$ \\
MKW8 & 220.16445 & 3.47031 & 0.0269 & 463.8 $_{-18.8} ^{+20.4}$ & 1.08 $_{-0.03} ^{+0.03}$ & 3.68 $_{-0.30} ^{+0.32}$ & 1.49 $_{-0.04} ^{+0.04}$ & 3.86 $_{-0.31} ^{+0.34}$ \\
A2029 & 227.73382 & 5.74455 & 0.0774 & 955.1 $_{-26.0} ^{+32.5}$ & 1.85 $_{-0.03} ^{+0.04}$ & 19.50 $_{-1.07} ^{+1.33}$ & 2.60 $_{-0.05} ^{+0.06}$ & 21.60 $_{-1.18} ^{+1.47}$ \\
A2052 & 229.18537 & 7.02162 & 0.0349 & 450.1 $_{-25.4} ^{+33.8}$ & 1.05 $_{-0.04} ^{+0.05}$ & 3.40 $_{-0.39} ^{+0.51}$ & 1.45 $_{-0.05} ^{+0.07}$ & 3.56 $_{-0.40} ^{+0.54}$ \\
A2061 & 230.30289 & 30.63350 & 0.0773 & 807.7 $_{-29.2} ^{+34.1}$ & 1.50 $_{-0.04} ^{+0.04}$ & 10.40 $_{-0.76} ^{+0.88}$ & 2.09 $_{-0.05} ^{+0.06}$ & 11.25 $_{-0.82} ^{+0.95}$ \\
MKW3s & 230.46593 & 7.70879 & 0.0448 & 549.3 $_{-26.2} ^{+35.3}$ & 1.09 $_{-0.03} ^{+0.05}$ & 3.87 $_{-0.37} ^{+0.50}$ & 1.51 $_{-0.05} ^{+0.06}$ & 4.06 $_{-0.39} ^{+0.52}$ \\
A2065 & 230.62280 & 27.70521 & 0.0735 & 1101.8 $_{-40.4} ^{+45.8}$ & 1.91 $_{-0.05} ^{+0.05}$ & 21.28 $_{-1.56} ^{+1.77}$ & 2.69 $_{-0.07} ^{+0.07}$ & 23.66 $_{-1.74} ^{+1.97}$ \\
A2063 & 230.77138 & 8.60957 & 0.0344 & 608.1 $_{-28.9} ^{+34.2}$ & 1.22 $_{-0.04} ^{+0.05}$ & 5.29 $_{-0.50} ^{+0.60}$ & 1.68 $_{-0.05} ^{+0.06}$ & 5.60 $_{-0.53} ^{+0.63}$ \\
A2107 & 234.91288 & 21.78285 & 0.0416 & 521.1 $_{-32.8} ^{+40.4}$ & 1.07 $_{-0.05} ^{+0.06}$ & 3.67 $_{-0.46} ^{+0.57}$ & 1.48 $_{-0.06} ^{+0.08}$ & 3.86 $_{-0.49} ^{+0.60}$ \\
A2142 & 239.58792 & 27.22996 & 0.0894 & 852.4 $_{-29.1} ^{+33.7}$ & 1.70 $_{-0.13} ^{+0.23}$ & 15.31 $_{-3.57} ^{+6.27}$ & 2.38 $_{-0.19} ^{+0.33}$ & 16.78 $_{-3.91} ^{+6.87}$ \\
A2147 & 240.55862 & 15.97118 & 0.0365 & 1068.2 $_{-25.3} ^{+30.2}$ & 2.02 $_{-0.15} ^{+0.27}$ & 24.16 $_{-5.42} ^{+9.53}$ & 2.84 $_{-0.21} ^{+0.37}$ & 27.05 $_{-6.07} ^{+10.67}$ \\
A2151 & 241.14913 & 17.72143 & 0.0349 & 851.6 $_{-27.8} ^{+31.1}$ & 1.68 $_{-0.14} ^{+0.27}$ & 13.84 $_{-3.48} ^{+6.61}$ & 2.34 $_{-0.20} ^{+0.37}$ & 15.13 $_{-3.80} ^{+7.22}$ \\
AWM4 & 241.23602 & 23.93267 & 0.0320 & 315.7 $_{-23.1} ^{+33.4}$ & 0.81 $_{-0.04} ^{+0.06}$ & 1.58 $_{-0.23} ^{+0.34}$ & 1.12 $_{-0.05} ^{+0.08}$ & 1.63 $_{-0.24} ^{+0.35}$ \\
G049.33+44.38 & 245.12627 & 29.89331 & 0.0964 & 555.0 $_{-26.8} ^{+45.3}$ & 1.23 $_{-0.04} ^{+0.07}$ & 5.82 $_{-0.57} ^{+0.95}$ & 1.71 $_{-0.06} ^{+0.09}$ & 6.19 $_{-0.61} ^{+1.01}$ \\
A2199 & 247.15930 & 39.55093 & 0.0306 & 669.0 $_{-18.1} ^{+21.2}$ & 1.46 $_{-0.11} ^{+0.21}$ & 9.03 $_{-2.14} ^{+3.89}$ & 2.02 $_{-0.16} ^{+0.29}$ & 9.73 $_{-2.30} ^{+4.19}$ \\
A2244 & 255.67738 & 34.06060 & 0.1004 & 885.7 $_{-40.9} ^{+58.4}$ & 1.54 $_{-0.05} ^{+0.07}$ & 11.36 $_{-1.06} ^{+1.50}$ & 2.14 $_{-0.07} ^{+0.09}$ & 12.31 $_{-1.15} ^{+1.63}$ \\
A2249 & 257.44080 & 34.45566 & 0.0838 & 894.7 $_{-49.7} ^{+54.5}$ & 1.56 $_{-0.06} ^{+0.06}$ & 11.69 $_{-1.31} ^{+1.43}$ & 2.18 $_{-0.08} ^{+0.09}$ & 12.69 $_{-1.42} ^{+1.55}$ \\
A2255 & 258.18160 & 64.06303 & 0.0800 & 928.3 $_{-35.6} ^{+43.4}$ & 1.58 $_{-0.04} ^{+0.05}$ & 12.12 $_{-0.93} ^{+1.13}$ & 2.21 $_{-0.06} ^{+0.07}$ & 13.17 $_{-1.01} ^{+1.23}$ \\
NGC6338 & 258.84579 & 57.41119 & 0.0290 & 478.0 $_{-23.1} ^{+27.9}$ & 1.05 $_{-0.03} ^{+0.04}$ & 3.41 $_{-0.33} ^{+0.40}$ & 1.45 $_{-0.05} ^{+0.06}$ & 3.57 $_{-0.35} ^{+0.42}$ \\
bNGC6338 & 258.84653 & 57.43462 & 0.0291 & 482.1 $_{-22.8} ^{+28.0}$ & 1.06 $_{-0.03} ^{+0.04}$ & 3.45 $_{-0.33} ^{+0.40}$ & 1.46 $_{-0.05} ^{+0.06}$ & 3.62 $_{-0.34} ^{+0.42}$ \\
G345.40-39.34 & 312.98733 & -52.63006 & 0.0455 & 742.8 $_{-25.6} ^{+23.7}$ & 1.44 $_{-0.03} ^{+0.03}$ & 8.79 $_{-0.61} ^{+0.57}$ & 2.00 $_{-0.05} ^{+0.04}$ & 9.54 $_{-0.66} ^{+0.61}$ \\
bG345.40-39.34 & 312.99799 & -52.78586 & 0.0452 & 750.7 $_{-24.3} ^{+28.4}$ & 1.42 $_{-0.03} ^{+0.04}$ & 8.45 $_{-0.55} ^{+0.65}$ & 1.97 $_{-0.04} ^{+0.05}$ & 9.08 $_{-0.59} ^{+0.69}$ \\
A2457 & 338.92141 & 1.48650 & 0.0582 & 597.1 $_{-37.9} ^{+52.6}$ & 1.15 $_{-0.05} ^{+0.07}$ & 4.56 $_{-0.58} ^{+0.80}$ & 1.59 $_{-0.07} ^{+0.09}$ & 4.84 $_{-0.62} ^{+0.85}$ \\
A2572 & 349.30335 & 18.70286 & 0.0392 & 467.6 $_{-34.2} ^{+51.7}$ & 0.97 $_{-0.05} ^{+0.07}$ & 2.68 $_{-0.39} ^{+0.59}$ & 1.33 $_{-0.07} ^{+0.10}$ & 2.80 $_{-0.41} ^{+0.62}$ \\
A2593 & 351.08690 & 14.64490 & 0.0412 & 524.8 $_{-31.2} ^{+33.9}$ & 1.15 $_{-0.05} ^{+0.05}$ & 4.54 $_{-0.54} ^{+0.59}$ & 1.59 $_{-0.06} ^{+0.07}$ & 4.79 $_{-0.57} ^{+0.62}$ \\
A2597 & 351.33235 & -12.12388 & 0.0831 & 438.2 $_{-34.2} ^{+54.4}$ & 1.03 $_{-0.05} ^{+0.09}$ & 3.41 $_{-0.54} ^{+0.85}$ & 1.43 $_{-0.08} ^{+0.12}$ & 3.57 $_{-0.57} ^{+0.90}$ \\
A2626 & 354.12631 & 21.14673 & 0.0558 & 650.4 $_{-52.0} ^{+66.5}$ & 1.15 $_{-0.06} ^{+0.08}$ & 4.60 $_{-0.74} ^{+0.94}$ & 1.61 $_{-0.09} ^{+0.11}$ & 4.96 $_{-0.80} ^{+1.02}$ \\
A4038 & 356.92869 & -28.14290 & 0.0298 & 743.7 $_{-27.6} ^{+30.4}$ & 1.25 $_{-0.03} ^{+0.03}$ & 5.78 $_{-0.43} ^{+0.47}$ & 1.74 $_{-0.04} ^{+0.05}$ & 6.13 $_{-0.46} ^{+0.50}$ \\
A2665 & 357.71104 & 6.14991 & 0.0565 & 564.6 $_{-46.0} ^{+63.8}$ & 0.84 $_{-0.05} ^{+0.06}$ & 1.80 $_{-0.30} ^{+0.42}$ & 1.20 $_{-0.07} ^{+0.09}$ & 2.07 $_{-0.34} ^{+0.48}$ \\
A4059 & 359.25423 & -34.75899 & 0.0494 & 611.0 $_{-34.1} ^{+42.9}$ & 0.83 $_{-0.03} ^{+0.04}$ & 1.72 $_{-0.19} ^{+0.24}$ & 1.14 $_{-0.04} ^{+0.05}$ & 1.78 $_{-0.20} ^{+0.25}$ \\

\hline

\end{tabular}
}


\end{minipage}

\end{table*}

\begin{table*}

\begin{minipage}{185mm}

\caption{Substructure measures, BCG coordinates, offsets and magnitude gaps for the 72 clusters. First we list the values of the three substructure tests from \citet{pin96} and their significance levels ($\beta$, $\Delta$ or DS and Lee 3D tests). Next we give the AD statistic and its associated $p-$value. Finally, we list the BCG coordinates, absolute magnitude, BCG offset to the X-ray center (in kpc and R$_{500}$) and the BCG velocity offset.}
\label{tab:sub_props}

{\scriptsize
\begin{tabular}{@{}lccccccccccccccc}

\hline

 name & $\beta$ & $\beta_{sig}$ & $\Delta$ & $\Delta_{s}$ & L3D & L3D$_{s}$ & AD & $p_{AD}$ & ra[BCG] & dec[BCG] & M$_r$[BCG] & $\Delta$m$_{12}$ & Offset & Offset & $\Delta_{v}$ \\


 & & & & & & & & & (J2000) & (J2000) & & & (kpc) & (R$_{500}$) & (km s$^{-1}$) \\

\hline

A2734 & 1.6 & 0.202 & 83.4 & 0.008 & 1.2 & 0.866 & 0.5 & 2.6e-01 & 2.84021 & -28.85433 & -23.73 & 2.61 & 1.4 & 0.002 & 175.7 \\
A85 & 4.2 & 0.156 & 226.6 & 0.012 & 1.2 & 0.421 & 1.3 & 1.7e-03 & 10.46021 & -9.30318 & -23.73 & 1.11 & 2.3 & 0.002 & -30.0 \\
A119 & 24.0 & 0.012 & 324.6 & 0.000 & 1.1 & 0.543 & 0.5 & 2.1e-01 & 14.06715 & -1.25537 & -23.48 & 0.49 & 17.9 & 0.013 & 113.0 \\
A193 & -29.4 & 0.754 & 49.8 & 0.635 & 1.5 & 0.082 & 0.4 & 3.2e-01 & 21.28179 & 8.69923 & -22.73 & 0.60 & 2.8 & 0.003 & 120.5 \\
EXO0422 & -29.3 & 0.106 & 4.0 & 0.998 & 2.2 & 0.715 & 0.2 & 8.9e-01 & 66.46392 & -8.56069 & -22.15 & 0.11 & 3.8 & 0.007 & -302.8 \\
A496 & -9.5 & 0.926 & 278.1 & 0.054 & 1.2 & 0.012 & 0.2 & 7.6e-01 & 68.40768 & -13.26194 & -23.87 & 1.93 & 0.9 & 0.001 & 66.0 \\
S0540 & 3.1 & 0.116 & 16.5 & 0.036 & 2.0 & 0.257 & 0.2 & 9.4e-01 & 85.02775 & -40.83672 & -23.09 & 1.71 & 24.8 & 0.028 & -208.4 \\
A548e & 5.6 & 0.148 & 192.3 & 0.002 & 1.2 & 0.126 & 1.7 & 2.6e-04 & 87.15983 & -25.47786 & -23.13 & 0.16 & 0.5 & 0.000 & -212.0 \\
A3376 & -7.1 & 0.482 & 107.7 & 0.669 & 1.6 & 0.000 & 0.7 & 8.6e-02 & 90.54045 & -39.94987 & -22.74 & 0.69 & 322.4 & 0.287 & -152.0 \\
bA3376 & 14.7 & 0.086 & 91.9 & 0.451 & 1.5 & 0.000 & 0.8 & 4.5e-02 & 90.54045 & -39.94987 & -22.74 & 0.69 & 32.8 & 0.029 & -174.8 \\
A3391 & 10.0 & 0.188 & 80.3 & 0.168 & 1.2 & 0.489 & 0.4 & 3.2e-01 & 96.58415 & -53.69285 & -22.03 & -0.09 & 3.2 & 0.002 & -87.2 \\
A3395 & 5.1 & 0.160 & 187.2 & 0.086 & 1.4 & 0.004 & 0.5 & 2.1e-01 & 96.70658 & -54.54296 & -23.63 & 0.37 & 25.7 & 0.018 & 177.2 \\
bA3395 & 10.7 & 0.090 & 224.7 & 0.026 & 1.2 & 0.365 & 0.4 & 4.2e-01 & 96.90121 & -54.44948 & -23.26 & -0.37 & 11.6 & 0.008 & -586.7 \\
bA754 & 41.7 & 0.000 & 394.4 & 0.054 & 1.2 & 0.000 & 0.3 & 6.6e-01 & 137.13499 & -9.62988 & -23.78 & 1.15 & 242.2 & 0.154 & 137.9 \\
A754 & 42.8 & 0.000 & 334.0 & 0.118 & 1.3 & 0.000 & 0.2 & 7.9e-01 & 137.33005 & -9.69982 & -22.62 & 0.12 & 62.2 & 0.041 & -12.9 \\
G269.51+26.42 & 16.8 & 0.026 & 447.6 & 0.000 & 1.1 & 0.531 & 0.9 & 1.7e-02 & 159.17842 & -27.52833 & -22.10 & 0.32 & 6.7 & 0.006 & 104.3 \\
G241.85+51.53 & -9.6 & 0.422 & 88.7 & 0.018 & 1.3 & 0.383 & 0.4 & 3.0e-01 & 159.91133 & 5.17570 & -22.39 & -0.05 & 75.7 & 0.061 & -469.5 \\
USGCS152 & 4.2 & 0.382 & 10.8 & 0.497 & 2.2 & 0.204 & 0.2 & 8.6e-01 & 162.60871 & -12.84508 & -22.65 & 2.81 & 0.1 & 0.000 & -33.9 \\
G172.88+65.32 & -14.3 & 0.528 & 45.0 & 0.852 & 1.4 & 0.072 & 0.4 & 3.7e-01 & 167.93169 & 40.82076 & -23.52 & 0.67 & 117.5 & 0.093 & 703.9 \\
bG243.57+67.76 & -27.2 & 0.780 & 83.0 & 0.012 & 1.2 & 0.703 & 0.4 & 2.7e-01 & 173.21320 & 14.46118 & -23.45 & 0.57 & 31.7 & 0.023 & -101.3 \\
G243.57+67.76 & -21.9 & 0.682 & 80.2 & 0.026 & 1.3 & 0.477 & 0.4 & 3.2e-01 & 173.21320 & 14.46118 & -23.45 & 0.57 & 172.5 & 0.126 & -200.9 \\
G234.59+73.01 & 15.1 & 0.032 & 289.4 & 0.000 & 1.4 & 0.000 & 0.4 & 3.3e-01 & 176.00898 & 19.94983 & -22.62 & 0.34 & 468.9 & 0.381 & -229.0 \\
ZwCl1215 & -8.3 & 0.460 & 94.9 & 0.467 & 1.3 & 0.251 & 0.3 & 5.1e-01 & 184.42136 & 3.65585 & -23.04 & 0.38 & 2.1 & 0.001 & 2.4 \\
bA3528s & 46.1 & 0.002 & 178.7 & 0.028 & 1.4 & 0.004 & 1.4 & 1.3e-03 & 193.59267 & -29.01300 & -24.11 & 1.23 & 1.5 & 0.001 & 57.0 \\
A3528n & 46.1 & 0.000 & 178.7 & 0.030 & 1.4 & 0.004 & 1.4 & 1.3e-03 & 193.59267 & -29.01300 & -24.11 & 1.23 & 0.5 & 0.000 & 57.0 \\
A3528s & 77.9 & 0.000 & 161.2 & 0.154 & 1.7 & 0.000 & 1.1 & 6.7e-03 & 193.67087 & -29.22764 & -24.35 & 1.47 & 4.4 & 0.003 & 847.5 \\
A1644 & 0.2 & 0.276 & 320.0 & 0.012 & 1.3 & 0.000 & 0.4 & 4.1e-01 & 194.29829 & -17.40950 & -24.23 & 1.27 & 0.6 & 0.000 & 107.9 \\
A3532 & 49.8 & 0.002 & 120.2 & 0.038 & 1.3 & 0.094 & 0.4 & 3.8e-01 & 194.34154 & -30.36364 & -23.13 & 0.76 & 0.9 & 0.001 & -492.9 \\
A1650 & 10.0 & 0.112 & 129.0 & 0.798 & 1.4 & 0.002 & 0.4 & 2.8e-01 & 194.67289 & -1.76146 & -23.97 & 1.70 & 3.7 & 0.003 & 112.7 \\
A1651 & 10.8 & 0.118 & 110.3 & 0.695 & 1.2 & 0.194 & 0.6 & 9.1e-02 & 194.84354 & -4.19619 & -24.40 & 2.24 & 3.1 & 0.002 & 145.1 \\
G057.33+88.01 & 10.5 & 0.014 & 730.1 & 0.008 & 1.2 & 0.002 & 0.6 & 9.4e-02 & 194.89877 & 27.95926 & -23.07 & 0.12 & 75.2 & 0.045 & 210.5 \\
A1736 & 34.8 & 0.010 & 160.1 & 0.030 & 1.3 & 0.212 & 1.3 & 2.1e-03 & 201.86667 & -27.32478 & -23.88 & 0.58 & 531.1 & 0.383 & -56.1 \\
A3558 & 30.1 & 0.000 & 568.7 & 0.000 & 1.2 & 0.006 & 0.9 & 2.4e-02 & 201.98690 & -31.49553 & -24.64 & 1.29 & 0.4 & 0.000 & -179.0 \\
bA3558 & 62.0 & 0.000 & 683.2 & 0.000 & 1.4 & 0.000 & 2.3 & 7.1e-06 & 202.44896 & -31.60694 & -22.69 & -0.11 & 2.5 & 0.002 & 416.7 \\
bA3562 & 53.5 & 0.000 & 507.1 & 0.000 & 1.3 & 0.000 & 2.4 & 4.2e-06 & 202.86415 & -31.82069 & -23.09 & 0.24 & 3.4 & 0.002 & -548.0 \\
A3560 & -7.1 & 0.504 & 140.4 & 0.184 & 1.2 & 0.150 & 0.4 & 3.6e-01 & 203.10563 & -33.13781 & -22.90 & 0.75 & 33.4 & 0.029 & 75.8 \\
A3562 & 20.7 & 0.018 & 251.8 & 0.000 & 1.2 & 0.162 & 0.6 & 1.0e-01 & 203.39483 & -31.67236 & -24.07 & 1.81 & 1.9 & 0.001 & -4.2 \\
A1767 & 12.1 & 0.070 & 106.1 & 0.186 & 1.2 & 0.551 & 1.0 & 1.5e-02 & 204.03468 & 59.20640 & -23.57 & 1.00 & 31.7 & 0.022 & 120.2 \\
A1775 & -27.2 & 0.574 & 35.4 & 0.166 & 1.4 & 0.421 & 0.6 & 1.3e-01 & 205.45476 & 26.37347 & -23.87 & 1.78 & 8.5 & 0.009 & 98.9 \\
A3571 & 39.7 & 0.004 & 120.1 & 0.531 & 1.2 & 0.565 & 0.3 & 6.9e-01 & 206.86825 & -32.86497 & -24.36 & 2.63 & 4.1 & 0.003 & -211.7 \\
A1795 & 28.4 & 0.050 & 111.4 & 0.295 & 1.4 & 0.034 & 0.6 & 1.3e-01 & 207.21876 & 26.59293 & -23.66 & 1.93 & 5.3 & 0.004 & 100.4 \\
A1831 & -9.4 & 0.226 & 46.3 & 0.002 & 2.0 & 0.004 & 0.6 & 9.6e-02 & 209.78644 & 28.02260 & -23.17 & 0.51 & 231.4 & 0.253 & 432.9 \\
bMKW8 & -13.3 & 0.342 & 21.8 & 0.866 & 1.6 & 0.170 & 0.5 & 1.5e-01 & 219.43899 & 3.80707 & -21.31 & 0.60 & 403.5 & 0.442 & 481.8 \\
MKW8 & 1.1 & 0.378 & 102.2 & 0.467 & 1.2 & 0.279 & 0.4 & 3.6e-01 & 220.16261 & 3.46974 & -21.91 & -0.96 & 3.8 & 0.004 & -81.8 \\
A2029 & 7.3 & 0.194 & 188.3 & 0.008 & 1.2 & 0.068 & 1.0 & 1.1e-02 & 227.73375 & 5.74477 & -24.38 & 1.87 & 1.2 & 0.001 & 176.0 \\
A2052 & -43.2 & 0.998 & 87.8 & 0.585 & 1.3 & 0.244 & 0.3 & 7.0e-01 & 229.18536 & 7.02162 & -23.13 & 1.04 & 0.0 & 0.000 & -98.0 \\
A2061 & -21.0 & 0.904 & 145.0 & 0.162 & 1.4 & 0.004 & 1.1 & 7.3e-03 & 230.33576 & 30.67093 & -23.54 & 0.34 & 247.5 & 0.165 & 446.1 \\
MKW3s & 51.5 & 0.006 & 53.4 & 0.675 & 1.4 & 0.098 & 0.3 & 5.0e-01 & 230.46604 & 7.70882 & -22.73 & 0.40 & 0.4 & 0.000 & -189.8 \\
A2065 & -5.7 & 0.604 & 204.8 & 0.006 & 1.1 & 0.824 & 0.7 & 6.2e-02 & 230.60009 & 27.71437 & -22.60 & 0.01 & 112.3 & 0.059 & -1325.7 \\
A2063 & -16.4 & 0.828 & 126.2 & 0.068 & 1.2 & 0.212 & 0.7 & 7.1e-02 & 230.77209 & 8.60922 & -22.74 & 1.08 & 1.9 & 0.002 & -87.5 \\
A2107 & -16.5 & 0.726 & 128.3 & 0.016 & 1.1 & 0.868 & 0.6 & 9.9e-02 & 234.91269 & 21.78272 & -23.50 & 1.01 & 0.6 & 0.001 & 93.8 \\
A2142 & -5.1 & 0.490 & 167.3 & 0.172 & 1.2 & 0.070 & 0.5 & 2.0e-01 & 239.58334 & 27.23341 & -23.43 & -0.00 & 32.1 & 0.019 & 416.4 \\
A2147 & 53.4 & 0.000 & 547.4 & 0.000 & 1.3 & 0.000 & 1.0 & 1.5e-02 & 240.57095 & 15.97463 & -23.18 & 0.13 & 32.2 & 0.016 & -343.6 \\
A2151 & 63.2 & 0.000 & 552.4 & 0.000 & 1.2 & 0.044 & 0.3 & 6.6e-01 & 241.14915 & 17.72156 & -22.55 & 0.30 & 0.3 & 0.000 & 49.8 \\
AWM4 & 6.6 & 0.428 & 53.2 & 0.062 & 1.5 & 0.126 & 0.6 & 9.2e-02 & 241.23614 & 23.93266 & -23.20 & 1.55 & 0.3 & 0.000 & -21.3 \\
G049.33+44.38 & -31.2 & 0.562 & 31.1 & 0.810 & 1.4 & 0.357 & 0.3 & 6.4e-01 & 245.12970 & 29.89103 & -23.47 & 1.16 & 24.1 & 0.020 & -118.4 \\
A2199 & 7.0 & 0.182 & 268.0 & 0.253 & 1.3 & 0.000 & 0.4 & 4.6e-01 & 247.15933 & 39.55127 & -22.78 & 0.60 & 0.7 & 0.001 & 189.8 \\
A2244 & -17.4 & 0.580 & 77.0 & 0.064 & 1.1 & 0.938 & 1.0 & 1.4e-02 & 255.67705 & 34.05999 & -23.51 & 1.24 & 4.5 & 0.003 & -329.5 \\
A2249 & 31.9 & 0.006 & 115.8 & 0.010 & 1.3 & 0.084 & 0.5 & 2.6e-01 & 257.45277 & 34.45900 & -23.30 & 0.60 & 59.7 & 0.038 & 1073.0 \\
A2255 & 3.7 & 0.218 & 219.5 & 0.000 & 1.3 & 0.004 & 1.0 & 1.1e-02 & 258.14536 & 64.07070 & -23.23 & -0.18 & 95.8 & 0.061 & 728.8 \\
NGC6338 & 1.7 & 0.330 & 91.4 & 0.048 & 1.3 & 0.387 & 0.9 & 2.5e-02 & 258.84574 & 57.41119 & -22.96 & 0.87 & 0.1 & 0.000 & -516.8 \\
bNGC6338 & 1.7 & 0.314 & 91.4 & 0.054 & 1.3 & 0.345 & 0.9 & 2.5e-02 & 258.84574 & 57.41119 & -22.96 & 0.87 & 49.3 & 0.047 & -540.5 \\
G345.40-39.34 & -11.4 & 0.690 & 94.2 & 0.591 & 1.2 & 0.629 & 1.9 & 5.4e-05 & 312.98721 & -52.62978 & -23.68 & 0.82 & 1.2 & 0.001 & 487.5 \\
bG345.40-39.34 & -6.4 & 0.420 & 115.8 & 0.230 & 1.2 & 0.461 & 1.6 & 4.3e-04 & 312.98721 & -52.62978 & -23.68 & 0.82 & 505.7 & 0.357 & 564.5 \\
A2457 & -2.5 & 0.258 & 86.0 & 0.016 & 1.1 & 0.940 & 0.5 & 2.2e-01 & 338.91999 & 1.48489 & -23.49 & 0.99 & 8.7 & 0.008 & -60.0 \\
A2572 & 74.9 & 0.004 & 78.1 & 0.048 & 1.4 & 0.232 & 0.8 & 4.5e-02 & 349.30645 & 18.70814 & -22.76 & 0.64 & 17.1 & 0.018 & 293.8 \\
A2593 & 46.7 & 0.002 & 153.6 & 0.086 & 1.4 & 0.000 & 1.5 & 7.0e-04 & 351.08370 & 14.64713 & -23.23 & 0.81 & 11.3 & 0.010 & 127.7 \\
A2597 & -76.1 & 0.806 & 14.9 & 0.573 & 1.6 & 0.599 & 0.4 & 4.1e-01 & 351.33216 & -12.12410 & -23.13 & 2.22 & 1.7 & 0.002 & -19.8 \\
A2626 & -10.6 & 0.532 & 45.5 & 0.954 & 1.4 & 0.114 & 0.6 & 1.1e-01 & 354.12756 & 21.14735 & -23.67 & 1.09 & 5.2 & 0.004 & -362.4 \\
A4038 & 39.1 & 0.016 & 205.6 & 0.000 & 1.2 & 0.303 & 0.7 & 5.3e-02 & 356.93765 & -28.14075 & -23.75 & 1.22 & 17.6 & 0.014 & -287.2 \\
A2665 & -31.5 & 0.656 & 29.0 & 0.291 & 1.5 & 0.255 & 0.3 & 5.7e-01 & 357.71065 & 6.14960 & -23.65 & 2.53 & 2.0 & 0.002 & -115.4 \\
A4059 & -14.3 & 0.526 & 76.8 & 0.144 & 1.3 & 0.248 & 0.4 & 4.4e-01 & 359.25300 & -34.75914 & -24.20 & 2.39 & 3.6 & 0.004 & -115.4 \\

\hline

\end{tabular}
}


\end{minipage}

\end{table*}

\section{Optical galaxy distributions}

\begin{figure*}
\begin{center}
\leavevmode
\includegraphics[width=6.5in]{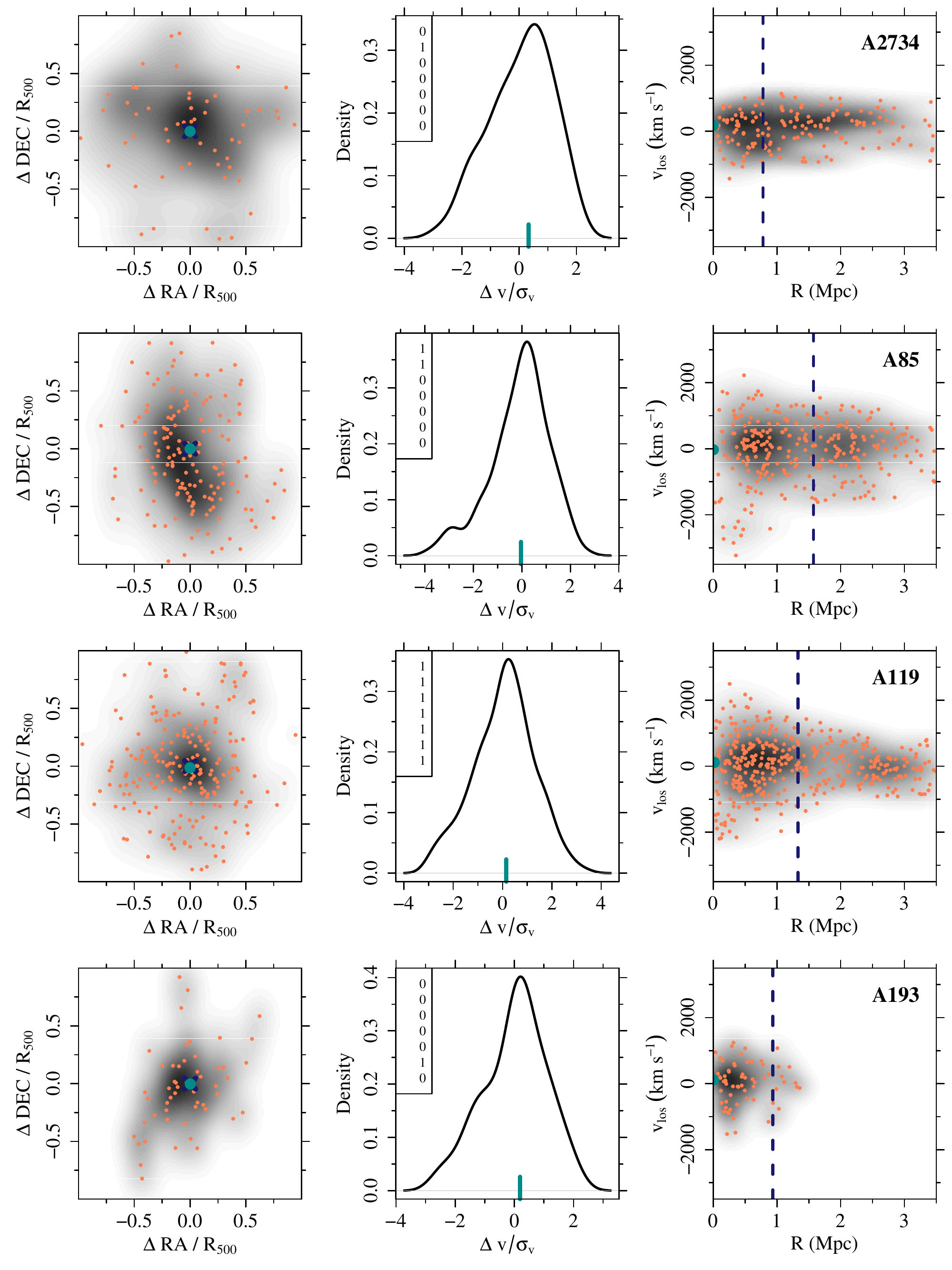}
\end{center}
\caption{Optical galaxy distributions for the 72 clusters in our sample. Each row shows the results for one cluster. In the left panels we display the projected sky distribution (in units of R$_{500}$). The blue cross indicates the X-ray center, while the cyan dot represents the BCG position (also in the right panels). In the central panels we show the velocity distributions, with the BCG indicated by the cyan bar. The left and central panels only have galaxies within R$_{500}$. The right panels display the phase-space diagrams. On all panels only members are displayed. The seven flags shown on the central panel mean 0 = relaxed and 1 = disturbed. The classifications (top down) are given by the AD, DS, BCG offset, C$_{SB}$, C$_{SB4}$, $\delta$ and n$_{core}$, respectively. Note: a portion of this figure is shown here for guidance regarding its form and content. A full version is available in the electronic edition of the MNRAS.}
\label{fig:opt_gals_dist}
\end{figure*}

Fig.~\ref{fig:opt_gals_dist} shows the optical galaxy distributions of the 72 systems used in this work. The projected sky distribution is on the left, while the velocity distribution is on the central panels and the phase-space is on the right. The left and center panels only have the member galaxies within R$_{500}$.

\section{Notes on individual clusters}

In the following we make specific remarks for some clusters of our sample.

\begin{enumerate}

\item {\bf Abell 1775}: There is a bright galaxy very close in projection to the BCG and the X-ray center. This galaxy could be mistakenly selected as the second brightest galaxy if a proper membership selection was not performed. The galaxy has a velocity offset of $-1793.4$ km s$^{-1}$. The cluster is classified as relaxed according to all the tests indicated in the central panel of Fig.~\ref{fig:opt_gals_dist}.

\item {\bf Abell 1831}: We think the redshift ($z = 0.0615$) reported in the literature is likely wrong. The reasons are explained below. There is a galaxy brighter than the chosen BCG and coincident to the X-ray position, but with velocity offset equal to $3868.2$ km s$^{-1}$. This is the BCG of the background cluster WHL J135915.1+275834, with photometric redshift $z_{phot} = 0.0746$ \citep{wen10}. Its BCG has redshift $z = 0.0759$. Hence, we have two clusters, Abell 1831 and WHL J135915.1+275834, nearly coincident on the plane of the sky but separated by $\sim 3900$ km s$^{-1}$. The X-ray emission is mostly associated to the background object. As the Abell catalog is based on a visual selection, most likely the chosen object had the brightest galaxy in the region. Note also that \citet{str99} mention that a background object was already indicated by \citet{che98}. In any case, for this work, we did the analysis around the original redshift listed to Abell 1831. The value we computed is $z = 0.0630$. Due to that, this cluster was classified as disturbed according to the BCG$-$X-ray offset. However, it is important to mention the galaxy distribution at $z \sim 0.0630$, within R$_{500}$, also indicates substructure from the DS and AD tests. On the contrary, all the X-ray parameters point to a relaxed cluster (see flags listed on Fig.~\ref{fig:opt_gals_dist}). As we mentioned above, the X-ray distribution is likely associated to the background object, that has a BCG coincident to the X-ray emission (indicating no substructure).

\item {\bf MKW8}: This cluster has two bright galaxies near the X-ray center.

\item {\bf bMKW8}: This object is nearly coincident with a background cluster with $z_{phhot} = 0.2193$ (WHL J143821.9+034013). ABELL 1942 is also close in projected coordinates. It is not clear if the X-ray emission is associated to the background system.

\item {\bf Abell 2142 (PLCKESZ G044.22+48.68)}: This cluster has two bright galaxies near the X-ray center.

\item {\bf bNGC6338}: This system is nearly identical to NGC6338, with very similar central coordinates, redshift and BCG.

\item {\bf Abell 754 (PLCKESZ G239.28+24.76)}: This cluster is very close on the plane of the sky to bA754 (bG239.28+24.76). In fact it is classified as an ongoing merger cluster \citep{ino16}. Both clusters show large offsets to the X-ray emission. It is very hard to separate the two systems and their galaxy components. As a consequence, the BCG of the first cluster is selected as the second brightest galaxy of the second cluster.

\item {\bf Abell 3376 (PLCKESZ G246.52-26.05)}: This system is also a double cluster, very close to bA3376 (bG246.52-26.05). It is classified as a dissociative merging cluster \citep{mon17}, with associated radio relics \citep{col17}. Both systems have large offsets to the X-ray emission.

\item {\bf Abell 3395 (PLCKESZ G263.20-25.21)}: This is another example of a double cluster, very close to bA3395 (bG2263.20-25.21). It was also previously classified as a merging system \citep{lak11}. In particular, according to \citet{don01} it appears to be nearly at first core passage. For the second cluster the BCG is coincident with the X-ray centroid.

\item {\bf Abell 3558 (PLCKESZ G311.99+30.71)}: This is one more double cluster, very close to bA3558 (bG311.99+30.71). This systems is located at the core of the Shapley supercluster \citep{han99}. It is also suggested as probably a merger seen just after the first core-core encounter \citet{bar02}. For both clusters the BCG is coincident with the respective X-ray position.

\item {\bf PLCKESZ G345.40-39.34 (Abell 3716S)}: This is another double cluster (bG345.40-39.34) in our sample. Differently than other objects listed above, the system is seen as double not only in the plane of the sky, but also in the phase space diagram. For the first cluster the BCG is coincident with the X-ray centroid. The two clusters are close enough in the plane of the sky, that the same BCG is chosen for the two systems (see Fig.~\ref{fig:opt_gals_dist}).

\item {\bf Abell 3562}: This a double cluster (bA3562), also part of the A3558 complex in the center of the Shapley supercluster \citep{bar02}. There is also a radio halo at the centre of the cluster A3562 \citep{ven03}. The BCGs of each cluster (A3562 and bA3562) are coincident to their respective X-ray peak, but the velocity offset of the BCG from bA3562 is $-548.0$ km s$^{-1}$. The inspection of the phase space of this cluster (see Fig.~\ref{fig:opt_gals_dist}) indicates the redshift distribution could be incomplete or perhaps we could have used the BCG as the cluster redshift. However, we preferred to follow the same procedure as for the other clusters on what regards the cluster redshift determination (see $\S$ \ref{cls_prop}).

\item {\bf Abell 3571 (PLCKESZ G316.34+28.54)}: This object is classified as relaxed by all our substructure tests (in the optical and X-rays), except for the $\beta$ test. Previous analysis also considered it as a relaxed system \citep{qui93, nev00}, but from a radio survey in the A3571 cluster complex \citet{ven02} suggest what is seen is the final stage of a merger event. The cluster A3571 is the final product after virialization of the merger, with gas and galaxy distributions relaxed within it, but unrelaxed in the outskirts. 

\item {\bf Abell 2151}: This object is classified as relaxed according to all X-ray substructure measurements, as well as the BCG offset and the AD test. But it is not according to the $\Delta$ test and the magnitude gap. From Fig.~\ref{fig:opt_gals_dist} we see there is a concentration of galaxies next to the BCG, within 0.5$\times$R$_{500}$, which explain the $\Delta$ test result and possibly the small magnitude gap between the first two BCGs.

\item {\bf Abell 4038}: This cluster is classified as disturbed by all optical tests, except the Lee 3D and $\Delta m_{12}$, but is considered as relaxed by all the four X-ray substructure parameters. The cluster has two bright galaxies in the center. A radio relic is also reported by \citet{sle98} and \citet{kal12}.

\end{enumerate}


\bsp	
\label{lastpage}
\end{document}